%% file: Primer.tex
\documentclass[12pt]{iopart}

\usepackage{epsfig,amsbsy,calc,graphics,latexsym,graphicx,psfrag,color,citesort}
\input{./Preamble/preamble}

\begin{document}

\title{A Primer for Manifestly Gauge Invariant Computations in $SU(N)$ Yang-Mills}

\author{
	Oliver J.~Rosten
}

\address{School of Physics and Astronomy,  University of Southampton,
	Highfield, Southampton SO17 1BJ, U.K.}
\ead{O.J.Rosten@soton.ac.uk}


\begin{abstract}
	It has recently been determined that, within the
	framework of the Exact Renormalization Group, continuum
	computations can be performed to any loop order
	in $SU(N)$ Yang-Mills theory without fixing the gauge or 
	specifying the details of the regularization scheme. In this paper,
	we summarise and refine the powerful diagrammatic techniques
	which facilitate this procedure and illustrate
	their application in the context of a calculation of
	the two-loop $\beta$ function.
\end{abstract}



\maketitle
\tableofcontents

\markboth{A Primer for Manifestly Gauge Invariant Computations in $SU(N)$ Yang-Mills}{A Primer for Manifestly Gauge Invariant Computations in $SU(N)$ Yang-Mills}

\section{Introduction and Conclusions}	\label{sec:Intro}

The Exact Renormalization Group (\ERG) has, for some
time now, provided a framework allowing continuum 
computations in $SU(N)$ Yang-Mills theory to be performed
without fixing the
gauge~\cite{GI-ERG-I,GI-ERG-II,YM-1-loop,Quarks2004,Thesis,YM-2-loop-A,YM-2-loop-B}. 
Whilst being of obvious
novelty value, manifest gauge invariance also provides
several technical benefits.
First,  the gauge field is protected from field strength 
renormalization. Secondly,
the Ward identities take a particularly simple form
since the Wilsonian effective action is built only from
gauge invariant combinations of the covariant derivative,
even at the quantum level~\cite{GI-ERG-I}. 
Thirdly, the difficult issue of
Gribov copies~\cite{Gribov}---which 
complicates non-perturbative studies in covariant
gauges---is entirely avoided.

This latter point highlights the point that a manifestly
gauge invariant formalism can, in many ways, be considered
naturally adapted for non-perturbative problems. Indeed,
the possibility of making statements about \eg confinement
in an entirely gauge independent manner is very appealing.
It is perhaps, then, something of a stroke of luck
that the manifestly gauge invariant scheme described in
this paper is formulated within the framework of the
ERG, since the ERG has a long and distinguished history
as a powerful tool for studying non-perturbative 
phenomena~\cite{Fisher:1998kv,TRM-elements,Aoki:2000wm,Litim:1998nf,Berges:2000ew,Bagnuls:2000ae,Polonyi:2001se,Salmhofer:2001tr,Delamotte:2003dw}.

The basic idea of the \ERG---the continuum version of
Wilson's RG~\cite{Wil,W&H,Pol}---is that of integrating
out degrees of freedom between the bare scale of a
quantum field theory and an effective scale, $\Lambda$.
The effects of these modes are encoded in the 
Wilsonian effective action, 
$S_\Lambda$, which describes the physics of the theory
in terms of parameters relevant to the effective scale.
Central to the \ERG\ methodology is the \ERG\ (or flow) equation
which determines
how the Wilsonian effective action
changes under infinitesimal changes of the scale.
It is by relating physics at different scales, in
this way, that the \ERG\ provides access to non-perturbative
phenomena.

There have been several different attempts to adapt
the ERG for non-Abelian gauge theory (for a comprehensive
review, see~\cite{Pawlowski:2005xe}). However, all
of these must face up to the problem that, at least
na\"ively, the implementation of a momentum cutoff
(which is fundamental to the \ERG) breaks non-Abelian
gauge invariance. The traditional solution to this
problem is to accept this breaking, recovering the 
physical symmetry in 
the limit that all quantum fluctuations
have been integrated 
out~\cite{Reuter:1993kw,Freire:2000bq,Litim:2002ce,Litim:2002hj,Bonini:1993sj,Bonini:1994kp,Ellwanger:1994iz,D'Attanasio:1996jd}.
In contrast, the scheme employed in this paper 
utilizes a regularization scheme based
on a real, gauge invariant cutoff, $\Lambda$~\cite{SU(N|N)}.
In this way (manifest) gauge invariance is maintained
at all scales.

Though the earliest
formulation of a manifestly gauge 
invariant \ERG~\cite{GI-ERG-I,GI-ERG-II} was constructed in the
large $N$-limit, its subsequent refinement~\cite{YM-1-loop}
facilitated the first manifestly gauge invariant calculation
of the one-loop $\beta$ function, $\beta_1$, at finite $N$. 
Guided by the universality of the answer, 
the calculation of $\beta_1$
was performed without specifying the precise details of the
regularization scheme, leading to a partially diagrammatic
computational methodology. 

However, preserving
these diagrammatic techniques beyond one loop 
required generalization of the flow equation.
The reason for this is as follows. The implementation
of the
gauge invariant cutoff is 
achieved by embedding
the physical $SU(N)$ theory in a spontaneously broken $SU(N|N)$
supergauge theory~\cite{SU(N|N)}. Besides the coupling,
$g(\Lambda)$, of the physical $SU(N)$ gauge field, $A^1_\mu$,
there is a second coupling, $g_2(\Lambda)$, associated with
an unphysical $SU(N)$ field, $A^2_\mu$, that requires separate 
renormalization~\cite{Quarks2004,Thesis,YM-2-loop-A}.\footnote{
For technical reasons, 
a superscalar field is given
zero mass dimension~\cite{YM-1-loop}, and thus is associated by the
usual dimensional reasoning with an infinite number of
dimensionless couplings. These couplings do not
require renormalization~\cite{Thesis,YM-2-loop-B}.} 
As a direct consequence of this, the diagrammatic
tricks can only be maintained if the flow equation
treats $A^1$ and $A^2$ independently, in the broken phase.
This 
challenging problem, finally overcome
in~\cite{Thesis,YM-2-loop-A}, yielded
the first manifestly gauge invariant, continuum
calculation of the two-loop  $\beta$ function~\cite{Thesis,YM-2-loop-B}, $\beta_2$.
The practical execution of this calculation
necessitated the major development
of the original diagrammatic techniques,
not only for calculational convenience, but also to elucidate
the structure of the new or flow
equation~\cite{Thesis,YM-2-loop-A,YM-2-loop-B}. 

The considerable amount of detail present in~\cite{Thesis,YM-2-loop-A,YM-2-loop-B}
reflects the subtlety and complexity of the new
flow equation. However, the actual
rules for performing perturbative calculations 
are remarkably simple, as a consequence
of the diagrammatics. In this paper, we summarize the 
primary diagrammatic rules
and use the illustration of their application
to significantly refine the calculational procedure of~\cite{YM-2-loop-A}
(see also~\cite{Thesis}). Though performing an actual
$\beta$-function calculation
is still more complicated than in alternative
approaches, the developments of this paper
allow a further radical simplification
which we will discuss later.

Contrary to previous works~\cite{Thesis,Quarks2004,YM-2-loop-A,YM-1-loop},
the flow equation is introduced via its diagrammatic representation.
Hence, we do not work with a single flow equation but rather with an 
infinite class which obey the same diagrammatic rules.
(It is assumed that the general properties which all good
flow equations possess~\cite{YM-1-loop,Quarks2004,jose,Scalar-1-loop,Scalar-2-loop} 
\eg invariance of physics under the flow and \etc are 
implicitly satisfied.)

Within our \ERG,
the flow is controlled by a (generically) non-universal object,
$\hS$, the `seed
action'~\cite{Thesis,YM-2-loop-A,YM-2-loop-B,Scalar-1-loop,Scalar-2-loop,YM-1-loop,giqed}. 
This respects the same symmetries as the Wilsonian effective 
action, $S$, and has the same structure. However, whereas our
aim is to solve the flow for $S$, $\hS$ acts as an input. In
accord with our general philosophy, 
the seed action is left unspecified where possible
and implicitly defined where necessary. There is one important
exception to this, at the heart of the diagrammatic
techniques: it is technically useful
to set the two-point, tree level seed action vertices equal to 
their Wilsonian effective action counterparts. In turn, this
ensures that, if the flow equation
is sufficiently general~\cite{YM-2-loop-A}, then 
for each independent two-point, tree level vertex (that cannot be consistently
set to zero) there exists
an `effective propagator', which plays a crucial
diagrammatic \role. These effective propagators, denoted
by $\Delta$, are the
inverses of the two-point, tree level vertices up to remainder
terms~\cite{YM-1-loop} (in the gauge sector); we call this
the `effective propagator
relation'~\cite{YM-1-loop}. The remainder terms appear as 
a consequence of the manifest gauge invariance: the effective
propagators are inverses of the two-point, tree level vertices
only in the transverse space.
It is important to emphasise that the effective propagators are
by no means propagators in the usual sense, but their name
recognizes their similarity in both form and diagrammatic function.

In this paper we will indicate in some detail 
how, starting from
a diagrammatic expression for $\beta_2$ involving
the seed action action and details of the covariantization
of the cutoff, we can derive an expression with no explicit
dependence on these non-universal objects. 
The basic strategy is to recognize that amongst the
terms generated by the flow equation are manipulable diagrams
comprising
exclusively Wilsonian effective action vertices joined
together by a differentiated effective propagator. These latter
objects are denoted by $\dd$ where, having defined
\be
\label{eq:alpha}
	\alpha \equiv \frac{g_2^2}{g^2},
\ee
we define
\be
\label{eq:dot}
	\bigdot{X} \equiv -\flowConstAl X;
\ee
$\flow$ being the generator of the \ERG\ flow. 
The manipulable diagrams are processed by moving
$\flowConstAl$ from the effective propagator
to strike the diagram as a whole, minus correction
terms in which $\flowConstAl$ strikes the vertices.
The former terms, called $\Lambda$-derivative terms,
are those from which the numerical value of $\beta_2$
can be directly extracted~\cite{Thesis,YM-2-loop-B};
the latter terms can be processed using
the flow equation and the resulting set of
diagrams simplified, using primary diagrammatic
identities. At this point, we are able to identify
cancellations of non-universal contributions, at the
diagrammatic level.
There is, however, a complication to this diagrammatic procedure:
particular classes of sub-diagrams can have two distinct diagrammatic
representations. The equivalence of these representations
is encoded in the secondary diagrammatic identities, of
which a sub-set are presented here.
(For the complete set, the reader is referred to~\cite{mgiuc}.)

Iterating the above procedure, the diagrammatic expression
can be reduced exclusively to $\Lambda$-derivative terms
and `$\alpha$-terms'. These latter terms explicitly
involve the flow of $\alpha$ and are an artefact of 
the $SU(N|N)$ regularization scheme. Subject to very
general constraints, they 
vanish, as they must, in the limit that 
$\alpha \rightarrow 0$~\cite{Thesis,YM-2-loop-B}.

Compared with~\cite{YM-2-loop-A},
we realize that, at each stage of the calculation,
both the cancellation of sets of non-universal contributions and
the creation of sets of $\Lambda$-derivative terms
can each be done in parallel. This vastly simplifies the calculational
procedure, the benefits becoming increasingly pronounced with
each loop order. Despite this important development,
even at two-loops a $\beta$ function calculation still has many steps.
However, the illustration of
the $\beta_2$ diagrammatics will serve to demonstrate
that the procedure is algorithmic.
As we will see in~\cite{RG2005,mgiuc} (see also~\cite{Thesis}
for an incomplete discussion), this can  be
turned very much to our advantage:
using the techniques of this paper, we can jump straight
from the initial expression for a $\beta$ function
coefficient  to the $\Lambda$-derivative and $\alpha$-terms, \emph{to all
orders in perturbation theory}.
At a stroke, this removes the major
difficulty associated with performing
computations within our framework, leaving open
the exciting prospect of a manageable, manifestly
gauge invariant calculus for $SU(N)$ Yang-Mills theory.

Section~\ref{sec:Elements} introduces the necessary
elements of $SU(N|N)$ gauge theory, for our purposes.
In section~\ref{sec:Diagrammatics} we describe the
diagrammatics. First, we give the diagrammatic
representation of the exact flow equation. Secondly, we
describe the diagrammatic realization of the Ward identities.
Thirdly, we show how the various vertices of the
objects involved in the flow equation can be Taylor expanded.
Lastly, we specialize the diagrammatics
to the perturbative domain.
In section~\ref{sec:Illustration}, we illustrate the
use of the diagrammatics in the context of a computation of
the perturbative two-loop $\beta$ function.

\section{Elements of $SU(N|N)$ Gauge Theory}	\label{sec:Elements}

Throughout this paper, we work in Euclidean dimension, D.
We regularize $SU(N)$ Yang-Mills by embedding
it in spontaneously broken $SU(N|N)$ Yang-Mills,
which is itself regularized by covariant higher derivatives~\cite{SU(N|N)}.
The supergauge field, $\SF_\mu$, is valued in
the Lie superalgebra and, using the defining representation,
can be written as a Hermitian supertraceless supermatrix:
\[
	\SF_\mu = 
	\left(
		\begin{array}{cc}
			A_\mu^1 	& B_\mu 
		\\
			\af{B}_\mu & A_\mu^2
		\end{array} 
	\right) + \SF_\mu^0 \one.
\]
Here, $A^1_\mu(x)\equiv A^1_{a\mu}\tau^a_1$ is the
physical $SU(N)$ gauge field, $\tau^a_1$ being the $SU(N)$
generators orthonormalized to
$\tr(\tau^a_1\tau^b_1)=\delta^{ab}/2$, while $A^2_\mu(x)\equiv
A^2_{a\mu}\tau^a_2$ is a second unphysical $SU(N)$ gauge field.
The $B$ fields are fermionic gauge fields which will gain a mass
of order $\Lambda$ from the spontaneous symmetry breaking; they play the
role of gauge invariant Pauli-Villars (PV) fields, furnishing the
necessary extra regularization to supplement the covariant
higher derivatives.

The theory is locally invariant under:
\be
\label{eq:GaugeT}
	\delta \SF_\mu = [\nabla_\mu,\Omega(x)] + \lambda_\mu \one.
\ee
The first term, in which $\nabla_\mu = \partial_\mu -i\SF_\mu$, 
generates supergauge transformations. Note that the coupling, $g$,
has been scaled out of this definition. It is worth doing
this: since we do not gauge fix, the exact preservation of~\eq{eq:GaugeT}
means that none of the fields suffer field strength 
renormalization, even in the broken phase~\cite{YM-1-loop}.

The second term in~\eq{eq:GaugeT} divides out the centre of the algebra.
This `no $\SF^0$ shift symmetry' ensures that nothing depends on $\SF^0_\mu$
and that $\SF^0_\mu$ has no degrees of freedom. We adopt a
prescription whereby we can effectively ignore the field $\SF^0_\mu$,  altogether, 
using it  to map us into a particular diagrammatic
picture~\cite{Thesis,YM-2-loop-A}.

For the superscalar field, $\SH$, which spontaneously
breaks the $SU(N|N)$ invariance, there is no
need to factor out the central term~\cite{SU(N|N)} and so we write
\[
\SH =
	\left(
		\begin{array}{cc}
			C^1		& D
		\\
			\af{D}	& C^2
		\end{array}
	\right).
\]
This field transforms homogeneously:
\[
	\delta \SH = -i [\SH,\Omega].
\]

In order that, at the classical level, the spontaneous breaking
scale tracks the covariant higher derivative effective cutoff
scale, $\Lambda$, we take $\SH$ to be dimensionless and demand
that $\hS$ has the minimum of its effective potential at
\be
\label{sigma}
<\SH>\ = \sigma \equiv \pmatrix{\one & 0\cr 0 & -\one}.
\ee
In this case the classical action $S_0$ also has a minimum 
at~\eq{sigma}. 
Ensuring that this is not destroyed by quantum corrections demands
that the Wilsonian effective action one-point $C^1$, $C^2$ vertices
vanish~\cite{YM-1-loop,Thesis,YM-2-loop-A}, which can be translated
into a constraint on $\hS$. 

Working in the broken phase, the fermionic fields $B_\mu,\af{B}_\nu$ and
$D,\af{D}$ can be combined into the fields, 
\numparts
\bea
\label{eq:F}
	F_M & = & (B_\mu, D),
\\
\label{eq:Fbar}
	\bar{F}_N & = & (\bar{B}_\nu, -\bar{D}),
\eea
\endnumparts
where $M$, $N$ are
five-indices~\cite{Thesis,YM-2-loop-A,YM-2-loop-B}.\footnote{The summation
convention for these indices is that we take each product of
components to contribute with unit weight.} 
This simplification recognizes that,
via the Higgs mechanism, $B$ and $D$ gauge transform into each other
and so propagate together.

In $SU(N|N)$ gauge theory, the supertrace replaces
the trace as the natural cyclic invariant~\cite{SU(N|N),Bars}. 
The manifestly gauge invariant
Wilsonian effective action, $S$, 
comprises supertraces and products of supertraces where
the arguments of the supertraces are sets
of net-bosonic fields.

\section{Diagrammatics}	\label{sec:Diagrammatics}

In recognition of the central \role\ played
by the diagrammatics, our approach is generally
to first state the diagrammatic rules and then to
describe the various elements involved.

\subsection{The Exact Flow Equation}

The \ERG\ equation can
be represented as shown in figure~\ref{fig:Flow}~\cite{Thesis,YM-2-loop-A}.
\bcf[h]
	\beas
	-\flow 
	\dec{
		\ensuremath{\begin{array}{c}\input{pstex/Vertex-S.pstex_t} \end{array}}
	}{\{f\}}
	& = & a_0[S,\Sigma_g]^{\{f\}} - a_1[\Sigma_g]^{\{f\}}
	\\
	& = &
	\frac{1}{2}
	\dec{
		\ensuremath{\begin{array}{c}\input{pstex/Dumbbell-S-Sigma_g.pstex_t} \end{array}} - \ensuremath{\begin{array}{c}\input{pstex/Padlock-Sigma_g.pstex_t} \end{array}} - \ensuremath{\begin{array}{c}\input{pstex/WBT-Sigma_g.pstex_t} \end{array}}
	}{\{f\}}
	\eeas
\caption{The diagrammatic form of the flow equation.}
\label{fig:Flow}
\ecf

The \lhs\ depicts the flow of all independent Wilsonian effective action
vertex \emph{coefficient functions}, 
which correspond to the set of
fields, $\{f\}$. Each coefficient function has associated
with it an implied supertrace structure (and symmetry factor which,
as one would want, does not appear in the diagrammatics).
For example,
\be
\label{eq:Vertex-C1C1}
	\dec{
		\ensuremath{\begin{array}{c}\input{pstex/Vertex-S.pstex_t} \end{array}}
	}{C^1C^1}
\ee
represents both the coefficient functions $S^{C^1 C^1}$ and
$S^{C^1,C^1}$ which, respectively, are associated with the
supertrace structures $\str C^1 C^1$ and $\str C^1 \str C^1$. 

The first diagram on the \rhs\ of figure~\ref{fig:Flow}
is a formed by the bilinear functional $a_0[S,\Sigma_g]$,
whereas the next two diagrams are formed by $a_1[\Sigma_g]$.
All three diagrams
have two different components. The lobes
represent vertices of action functionals,
where $\Sigma_g \equiv g^2S - 2 \hat{S}$. 
The object attaching
to the various lobes, \DummyKernel,  is
the sum over vertices of the covariantized \ERG\ kernels~\cite{GI-ERG-I,YM-1-loop}
and, like the action vertices, can be decorated by fields belonging to $\{f\}$.
The fields of the action vertex (vertices) to which the vertices of the kernels attach
act as labels for the \ERG\ kernels
though, in certain circumstances, the particular decorations of
the kernel are required for unambiguous identification~\cite{Thesis,YM-2-loop-A}.
However, in actual calculations, these non-universal details are irrelevant.
We loosely refer to both individual and summed over 
vertices of the kernels simply as a kernel. 
Note that kernels labelled at one end by either $A$ or $B$ and at the other
by either $C$ or $D$
do not exist~\cite{YM-1-loop}.

The final diagram on the \rhs\ contains a kernel
which `bites its own tail'. Such diagrams are not properly 
UV regularized by the $SU(N|N)$ regularization 
and, in the past, it has been argued that
they can be discarded~\cite{YM-2-loop-A,TRM-Faro,GI-ERG-I,YM-1-loop}.\footnote{
These diagrams are artefacts of the flow equation. The $SU(N)$
gauge theory \emph{is} fully regularized by the $SU(N|N)$ scheme. However,
regularization of the flow equation does not trivially follow from
the regularization of the underlying theory.}
Here, though, we will keep these diagrams: as recognized in~\cite{Thesis},
in any calculation of universal quantities, 
all explicit instances of diagrams in which a kernel
bites its tail  (which can always be dimensionally regularized)
are cancelled by implicit instances buried in other
terms. We will see an
example of this in section~\ref{sec:Illustration}.

At this point, is is worth drawing attention to a subtlety
of the $SU(N|N)$ regularization scheme. For the scheme
to be properly defined, a preregularizer must be used~\cite{SU(N|N)}.
For convenience, this has traditionally been taken to be dimensional
regularization. However, this amounts to a choice
which is by no means unique.  Indeed, as we shall see, there are strong
hints that there is an entirely diagrammatic prescription that
can be used instead, which would make sense in $D=4$.
Thus, using dimensional regularization to regularize 
diagrams in which the kernels bite their own tails is
distinct from its previous application as just a preregularizer.

The rule for decorating the complete diagrams on
the \rhs\ is simple: the set of fields, $\{f\}$, are distributed in 
all independent ways between the component objects of each diagram.

Embedded within the diagrammatic rules is a prescription for evaluating the
group theory factors. 
Suppose that we wish to focus on the flow of a particular
vertex coefficient function, which necessarily has a unique
supertrace structure. 
For example, we might be interested in just
the $S^{C^1 C^1}$ component of~\eq{eq:Vertex-C1C1}.

On the \rhs\ of the flow equation, we must
focus on the components of each diagram
with precisely the same 
supertrace structure as the \lhs,
noting that the kernel, like the vertices,
has multi-supertrace contributions (for more
details see~\cite{Thesis,YM-2-loop-A}).
In this more explicit diagrammatic picture,
the kernel is to be considered a double
sided object.
Thus, whilst the dumbbell like term of figure~\ref{fig:Flow}
has at least one associated supertrace, the next two diagrams
has  at least two, on a account of the loop
(this is strictly true only in the
case that kernel attaches to fields on the same
supertrace). If a closed
circuit formed by a kernel is devoid
of fields then
it contributes 
a factor of $\pm N$, depending on
the flavours of the fields to which the kernel forming
the loop attaches. This is most easily appreciated by
defining the projectors
\[
	\sigma_{+} \equiv
	\left(
		\begin{array}{cc}
			\one	&	0
		\\
				0	&	0
		\end{array}
	\right), \qquad
		\sigma_{-} \equiv
	\left(
		\begin{array}{cc}
			0	&	0
		\\
			0	&	\one
		\end{array}
	\right)
\]
and noting that $\str \sigma_\pm = \pm N$. 
In the counterclockwise sense, a $\sigma_+$
can always be inserted after an $A^1$, $C^1$ or $\bar{F}$,
whereas a $\sigma_-$
can always be inserted after an $A^2$, $C^2$ or $F$.

The rules thus described receive $1/N$ corrections in 
the $A^1$ and $A^2$ sectors. If a kernel
attaches to an $A^1$ or $A^2$, it comprises a direct
attachment and an indirect attachment, as shown 
in figure~\ref{fig:Attach} (see~\cite{Thesis,YM-2-loop-A} for more detail).
\bcf[h]
	\[
		\ensuremath{\begin{array}{c}\input{pstex/Direct.pstex_t} \end{array}} \rightarrow \left. \ensuremath{\begin{array}{c}\input{pstex/Direct.pstex_t} \end{array}} \right|_{\mbox{direct}} + \frac{1}{N} \left[ \ensuremath{\begin{array}{c}\input{pstex/Indirect-2.pstex_t} \end{array}} - \ensuremath{\begin{array}{c}\input{pstex/Indirect-1.pstex_t} \end{array}} \right]
	\]
\caption{The $1/N$ corrections to the group theory factors.}
\label{fig:Attach}
\ecf 

We can thus consider the diagram on the \lhs\ as having been unpackaged,
to give the terms on the \rhs. The dotted lines in the diagrams with indirect
attachments serve to remind us where the loose end of the kernel attaches
in the parent diagram.

\subsection{Ward Identities}		\label{sec:WIDs}

\subsubsection{Application to Isolated Vertices}

All vertices, whether they belong to either
of the actions or to the covariantized kernels
are subject to Ward identities. Due to the
manifest gauge invariance, these take a 
particularly simple form, as shown in 
figure~\ref{fig:WIDs}. This is our first example
of a primary diagrammatic identity.
\bcf[h]
	\[
		\ensuremath{\begin{array}{c}\input{pstex/WID-contract.pstex_t} \end{array}} = \ensuremath{\begin{array}{c}\input{pstex/WID-PF.pstex_t} \end{array}} + \ensuremath{\begin{array}{c}\input{pstex/WID-PFb.pstex_t} \end{array}} - \ensuremath{\begin{array}{c}\input{pstex/WID-PB.pstex_t} \end{array}} - \ensuremath{\begin{array}{c}\input{pstex/WID-PBb.pstex_t} \end{array}} +\cdots
	\]
\caption{The Ward identities.}
\label{fig:WIDs}
\ecf

On the \lhs, we contract a vertex with the momentum of
the field which carries $p$. This field---which we will
call the active field---can be either
$A^1_\rho$, $A^2_\rho$, $F_R$ or $\af{F}_R$. In the first two cases,
the open triangle $\GRk$ represents $p_\rho$ whereas, in the
latter two cases, it represents $p_R \equiv (p_\rho, 2)$~\cite{Thesis,YM-2-loop-A}.
(Given that we often sum over all possible fields, we can take the
Feynman rule for $\GRk$ in the $C$-sector to be null.)
In all cases,  $\GRk$ is independent of $\Lambda$ and $\alpha$, which is
encoded in the following primary diagrammatic identities:
\numparts
\bea
\label{eq:LdL-GRk-Pert-a}
	\hspace{0.8em} \stackrel{\bullet}{\GRk} & =  & 0,
\\
\label{eq:dalpha-GRk-a}
	\begin{array}{c}
		\Dal
	\\[-1.5ex]
		\GRk
	\end{array} & = & 0,
\eea
\endnumparts
where $ \Dal\ \equiv \partial / \partial{\alpha}$.

On the \rhs, we push the contracted momentum forward onto 
the field which directly follows the active field, in the counterclockwise
sense, and pull back (with a minus sign) onto
the field which directly precedes the active field. 
Since our diagrammatics is permutation symmetric, the struck field---which
we will call the target field---can
be either $X$, $Y$ or any of the undrawn fields represented
by the ellipsis.
Any field(s) besides the active field and the
target field will be called spectators.
Note that we can take
$X$ and / or $Y$ to represent the end of a kernel.
In this case, the struck field is determined to be unambiguously
on one side of the (double sided) kernel; 
the contributions
in which the struck field is on the other side are included
in the ellipsis. 
This highlights the point that allowing the active
field to strike another field necessarily involves a partial
specification of the supertrace structure: it must be the case that
the struck field either directly followed or preceded the active
field. In turn, this means that the Feynman rule for particular
choices of the active and target fields can be zero. For example,
an $F$ can follow, but never precede an $A^1_\mu$, and so the 
pull back of an $A^1_\mu$ onto an $F$ should be assigned a value
of zero. 
The momentum routing follows in obvious manner: for example,
in the first diagram on the \rhs, momentum $q+p$ now flows into
the vertex. In the case that the active field is fermionic,
the field pushed forward / pulled back onto is transformed
into its opposite statistic partner. There are some signs 
associated with this in the $C$ and $D$-sectors, which we
will not require here~\cite{Thesis,YM-2-loop-A}; for 
calculations of universal quantities, they are hidden
by the diagrammatics.

The half arrow which terminates the pushed forward / pulled back
active field is of no significance and can go on either side
of the active field line. It is necessary to
keep the active field line---even though the active field
is no longer part of the vertex---in order that
we can unambiguously deduce flavour changes 
and momentum routing, without reference to the parent diagram.

We illustrate the the application of
the Ward identities by considering contracting
$\GRk$ into the Wilsonian the effective action
two-point vertex:
\be
\label{eq:GR-TP}
	\ensuremath{\begin{array}{c}\input{pstex/GR-TP.pstex_t} \end{array}} =  \ensuremath{\begin{array}{c}\input{pstex/GR-TP-PF.pstex_t} \end{array}} - \ensuremath{\begin{array}{c}\input{pstex/GR-TP-PB.pstex_t} \end{array}}.
\ee
Given that $\GRk$ is null in the $C^i$ sector,
the fields decorating the two-point
vertex on the \rhs\ can be either both $A^i$s
or both fermionic. In the former case, 
\eq{eq:GR-TP} reads:
\[
	p_\mu S^{A^i A^i}_{\mu \ \, \nu}(p) = S^{A^i}_{\nu}(0) - S^{A^i}_{\nu}(0) = 0
\]
where we note that $S^{A^i}_{\nu}$ is in fact zero by itself,
as follows by both Lorentz invariance and gauge invariance.
In the latter case, \eq{eq:GR-TP} reads:
\[
	p_M S^{\bar{F} \; F}_{M N}(p) = S^{C^2}(0) - S^{C^1}(0),
\]
where we have used~\eq{eq:F} and have discarded
contributions which go like $S^{A^i}_{\nu}(0)$.
However, the $S^{C^i}(0)$ must vanish. This
follows from demanding
that the minimum of the superhiggs potential is
not shifted by quantum corrections~\cite{YM-1-loop}.
Therefore, we arrive at the diagrammatic identity
\be
\label{eq:D-ID-GR-TP-A}
	\ensuremath{\begin{array}{c}\input{pstex/GR-TP.pstex_t} \end{array}} = 0.
\ee

\subsubsection{Application to Complete Diagrams}

Consider the flow of a (Wilsonian effective
action) vertex which is contracted with
the momentum of one of its fields. Suppose that,
in a addition to the active field, the vertex is
decorated by the set of fields $\{f'\}$.
Referring back to
figure~\ref{fig:Flow}, the \lhs\ becomes the sum of
vertices decorated by $\{f'\}$ where, for each
diagram in the sum, one of the elements of $\{f'\}$
is either pushed forward or pulled back onto by
the active field.

The \rhs\ of figure~\ref{fig:Flow} now
comprises three different types of diagram. The active field
can either push forward or pull back
\begin{enumerate}
	\item	onto one of the elements of $\{f'\}$;

	\item	round an action vertex onto an internal field;

	\item	onto the end of a kernel. 
\end{enumerate}
An example of a diagrams of either the second or third type
is shown in figure~\ref{fig:GR-InternalField}.
\bcf[h]
	\[
	\ensuremath{\begin{array}{c}\input{pstex/Dumbbell-S-PF-W-Sigma_g.pstex_t} \end{array}}
	\]
\caption{Example of a diagram in which the active field
strikes an internal field.}
\label{fig:GR-InternalField}
\ecf

Recalling that, if the active field is fermionic, the flavour
of the struck field will change, attachment corrections
of the type shown in figure~\ref{fig:Attach} must be
worked out \emph{after} the action of the active
field, and according to whether the field attached \emph{to}
(rather than the field at the end of the kernel, which will
be of a different flavour) is in the $A^{1,2}$ sector~\cite{Thesis}.

By gauge invariance, it must be the case that the sum
over all diagrams of the second and third types
vanishes~\cite{YM-2-loop-A} (see~\cite{Thesis} for
an explicit demonstration of this). This follows
since we are simply computing the flow of
a vertex (albeit one in which one of the fields can
be though of as having been struck by a active field).
From figure~\ref{fig:Flow}, we know that there
cannot be any surviving contributions in which internal
fields are pushed forward / pulled back onto.\footnote{In
$\beta$ function calculations, where active
fields arise in a different context, diagrams in which
internal fields are pushed forward / pulled back onto
can survive.}

\subsection{Taylor Expansion of Vertices}	\label{sec:Taylor}

For the formalism to be properly defined,
it must be the case that all vertices
are Taylor expandable to all orders
in momenta~\cite{GI-ERG-I,GI-ERG-II,TRM-elements}.
For the purposes of this paper, we need only
the diagrammatic rules for a particular scenario.
Consider a vertex which is part
of a complete diagram, decorated by some set of internal
fields and by a single external $A^1$ (or $A^2$).
The diagrammatic representation for the zeroth order expansion
in the momentum of the external field is all that is required
and is shown in figure~\ref{fig:TaylorExpansion}~\cite{Thesis,YM-2-loop-A}; 
note the similarity to figure~\ref{fig:WIDs}.
\bcf[h]
	\[
		\ensuremath{\begin{array}{c}\input{pstex/Taylor-Parent.pstex_t} \end{array}} = \ensuremath{\begin{array}{c}\input{pstex/Taylor-PFa.pstex_t} \end{array}} + \ensuremath{\begin{array}{c}\input{pstex/Taylor-PFb.pstex_t} \end{array}} - \ensuremath{\begin{array}{c}\input{pstex/Taylor-PBa.pstex_t} \end{array}} - \ensuremath{\begin{array}{c}\input{pstex/Taylor-PBb.pstex_t} \end{array}} +\cdots
	\]
\caption{Diagrammatic representation of zeroth order Taylor expansion.}
\label{fig:TaylorExpansion}
\ecf

The interpretation of the diagrammatics is as follows. In the first diagram
on the \rhs, the vertex is differentiated \wrt\ the momentum carried
by the field $X$, whilst holding the momentum of the preceding field fixed
(if the preceding field carries zero momentum, it is effectively
transparent to the momentum derivative~\cite{YM-2-loop-A} 
and so we go in a clockwise sense
to the first field
which carries non-zero momentum to determine the momentum held constant).
Of course, using our current diagrammatic notation, this latter field can be any of those
which decorate the vertex, and so we sum over all possibilities. 
Thus,
each cyclically ordered push forward like term has a partner,
cyclically ordered pull back like term, such that
the pair can be interpreted as
\be
	\left( \left. \partial^r_\mu \right|_s - \left. \partial^s_\mu \right|_r \right) \mathrm{Vertex},
\label{eq:Momderivs}
\ee
where $r$ and $s$ are momenta entering the vertex. In the case that $r=-s$, we can
and will
drop either the push forward like term or pull back like term, since
the combination can be expressed as $\partial^r_\mu$; we
interpret the diagrammatic notation appropriately.
Just as in figure~\ref{fig:WIDs}, 
the fields $X$ and or $Y$ can be interpreted as the end
of a kernel. In this case, we introduce some new
notation, since it proves confusing in complete diagrams
to actually locate the derivative
symbol at the end of such an object. The notation for the
derivative \wrt\ the momentum entering the end of a kernel 
is introduced in figure~\ref{fig:Taylor-EndofKernel}.
\bcf[h]
	\[
		\ensuremath{\begin{array}{c}\input{pstex/DifferentiatedKernel-A.pstex_t} \end{array}}
	\]
\caption{Notation for the derivative \wrt\ the momentum entering an
undecorated kernel.}
\label{fig:Taylor-EndofKernel}
\ecf

Recalling that a 
kernel, whose fields are explicitly
cyclically ordered, is a two-sided object, we first note
that the field whose momentum we have expanded in is sat on the
top-side of the vertex.
The derivative is taken to be \wrt\
the momentum which flows \emph{into} the end of the vertex which
follows the derivative, in the sense indicated by the
arrow on the derivative symbol.
It is clear that the direction of the
arrow on the derivative symbol can be reversed 
at the expense of a minus sign.

\subsection{Charge Conjugation Invariance} \label{sec:CC}

Charge conjugation invariance can be used to simplify the
diagrammatics, by allowing us to discard certain terms
and to combine others. The diagrammatic rule for replacing
a diagram with its charge conjugate is to reflect
the diagram, picking up a sign for every external $A^1$ or $A^2$
and letting $\bar{F} \leftrightarrow F$~\cite{Thesis,YM-2-loop-A}
(we temporarily assume that no Taylor expansions
have been performed and that, should any of the fields
be contracted with their momentum, the Ward identities are
yet to be applied).

Charge conjugation is of particular use in complete diagrams
for which all external fields are bosonic. In this case,
charge conjugation effectively never changes field flavours.
The only such changes induced by charge conjugation are in
 the fermionic sector where,
after reflection, we must change
$\bar{F} \leftrightarrow F$. However, by taking all
external fields to be bosonic, these changes
now only affect internal fields, whose flavours are
summed over anyway. 

Thus, for example, a diagram possessing only bosonic
external fields (and any number of internal fields)
must possess an even number of external $A^{1,2}$ fields
but any number of external $C^{1,2}$ fields.

It is straightforward to extend the diagrammatic rule
for charge conjugation
to include diagrams containing momentum derivatives
and / or application of the Ward identities~\cite{Thesis,YM-2-loop-A}. 
Supplementing
the previous rule, we simply pick up a minus sign for
each momentum derivative and for each application of
the Ward identities~\cite{Thesis,YM-2-loop-A}. Note that
active fields which have been processed by the Ward identities
should still be counted when we sum up the number of
external $A^1$s and $A^2$s; this is intuitive from a
diagrammatic point of view, since the field line is kept,
but now terminates in a half arrow, rather then entering
a vertex.

This allows us to simplify the set of terms generated
either by an application of the Ward identities or
by a Taylor expansion. To illustrate this, we need
deal only the former case, due to the similarity
of the diagrammatic rules for each.
Consider a diagram generated by a single application of
the Ward identities, in which no Taylor expansions have been
performed. 
We focus on a single target field, which we
know can be both pushed forward and pulled back onto,
as shown in figure~\ref{fig:CC-Combine-Ex}.
\bcf[h]
	\[
	\ensuremath{\begin{array}{c}\input{pstex/Stub-PF.pstex_t} \end{array}} - \ensuremath{\begin{array}{c}\input{pstex/Stub-PB.pstex_t} \end{array}}
	\]
\caption{Sum of the push forward and pull back onto the same target field.}
\label{fig:CC-Combine-Ex}
\ecf

The assumption that the diagram contains no further
target fields or momentum derivatives  tells us that the supertrace
structure of all spectator fields is unspecified. That all external
fields are bosonic ensures that charge conjugation leaves
the field content of the diagram unchanged.
Thus, reflecting either diagram of figure~\ref{fig:CC-Combine-Ex},
we can combine terms with the other diagram. Whether contributions
add or cancel depends on whether the diagrams as a whole are
charge conjugation odd or even.

Given that we have combined terms in this way, suppose that
there is a second active field (or that we perform a Taylor
expansion). Using the Ward identities, we once again find
that each target field is both pushed forward and pulled 
back onto. Now, however, such terms cannot be combined, since
the supertrace structure of the fields which spectate 
\wrt\ this second application of the Ward identities have
a partially specified supertrace structure, themselves.
Consider
the complete set of diagrams generated by any number of
applications of the Ward identities and possessing any
number of momentum derivatives. 
For each factorizable sub-diagram, we can combine
one push forward with one pull back;
thereafter, we cannot combine terms further.

In our example calculation of $\beta_2$, we will encounter
active fields attached to internal lines. Since the flavour
of internal lines is summed over, we can combine pushes forward
with pulls back of these fields, using the recipe above.

\subsection{Perturbative Diagrammatics}

In the perturbative domain, we have the following
weak coupling expansions~\cite{GI-ERG-I,YM-1-loop,Thesis,YM-2-loop-A}.
The Wilsonian effective action is given by
\be
	S = \sum_{i=0}^\infty \left( g^2 \right)^{i-1} S_i = \frac{1}{g^2}S_0 + S_1 + \cdots,
\label{eq:Weak-S}
\ee
where $S_0$ is the classical effective action and the $S_{i>0}$
the $i$th-loop corrections. The seed action has a similar expansion:
\be
	\hat{S} = \sum_{i=0}^\infty  g^{2i}\hat{S}_i.
\label{eq:Weak-hS}
\ee
Recalling~\eq{eq:alpha} we have:
\bea
	\beta \equiv \flow g 		& = & \sum_{i=1}^\infty  g^{2i+1} \beta_i(\alpha)
\label{eq:beta}
\\[1ex]
	\gamma \equiv \flow \alpha 	& = & \sum_{i=1}^{\infty}  g^{2i} \gamma_i(\alpha).
\label{eq:gamma}
\eea

\subsubsection{The Weak Coupling Flow Equations}

Defining $\Sigma_i = S_i - 2\hS_i$, the weak coupling flow equations
follow from substituting~\eq{eq:Weak-S}--\eq{eq:gamma}
into the flow equation, as shown in equation~\eq{eq:WeakFlow}~\cite{Thesis,YM-2-loop-A}.
	\be
		\dec{
			\ensuremath{\begin{array}{c}\input{pstex/Vertex-n-LdL.pstex_t} \end{array}} 
		}{\{f\}}
		= 
		\dec{
			\begin{array}{c}
				\ds
				\sum_{r=1}^n \left[2\left(n_r -1 \right) \beta_r +\gamma_r \pder{}{\alpha} \right]\ensuremath{\begin{array}{c}\input{pstex/Vertex-n_r-B.pstex_t} \end{array}} 
			\\[4ex]
				\ds
				+ \frac{1}{2} 
				\left( 
					\sum_{r=0}^n \ensuremath{\begin{array}{c}\input{pstex/Dumbbell-n_r-r.pstex_t} \end{array}} - \ensuremath{\begin{array}{c}\input{pstex/Vertex-Sigma_n_-B.pstex_t} \end{array}} - \ensuremath{\begin{array}{c}\input{pstex/WBT-Sigma_n-.pstex_t} \end{array}}
				\right)
			\end{array}
		}{\{f\}}
	\label{eq:WeakFlow}
	\ee

We refer to the first two terms on the \rhs\ of~\eq{eq:WeakFlow} as
$\beta$ and $\alpha$-terms, respectively.
The symbol $\bullet$, as in equation~\eq{eq:dot}, means
$-\flowConstAl$. A vertex whose argument is an unadorned letter, say $n$,
represents $S_n$. We define $n_r \equiv n-r$ and $n_\pm \equiv n \pm 1$. The
bar notation of the dumbbell term is defined as follows:
\[
	a_0[\bar{S}_{n-r}, \bar{S}_r] 	\equiv 	a_0[S_{n-r}, S_r] - a_0[S_{n-r}, \hat{S}_r] - a_0[\hat{S}_{n-r}, S_r].
\]

\subsubsection{The Effective Propagator Relation}	\label{sec:EPReln}

The effective propagator relation~\cite{YM-1-loop} is central
to the perturbative diagrammatic approach, and arises
from examining the flow of all two-point, tree level vertices.
This is done by setting $n=0$ in~\eq{eq:WeakFlow}
and specializing $\{f\}$ to contain two fields, 
as shown in equation~\eq{eq:TLTP-flow}.
We note that we can and do choose
all such vertices to be single supertrace terms~\cite{Thesis,YM-2-loop-A}.
	\be
		\ensuremath{\begin{array}{c}\input{pstex/Vertex-TLTP-LdL.pstex_t} \end{array}} = \ensuremath{\begin{array}{c}\input{pstex/Dumbbell-S_0-Sigma_0.pstex_t} \end{array}}
	\label{eq:TLTP-flow}
	\ee

Following~\cite{TRM-Faro,GI-ERG-I,GI-ERG-II,YM-1-loop,Thesis,YM-2-loop-A,Scalar-2-loop},
we use the freedom inherent in $\hat{S}$ by choosing the two-point, tree
level seed action vertices equal to the corresponding Wilsonian effective
action vertices. Equation~\eq{eq:TLTP-flow} now simplifies.
Rearranging, integrating \wrt\ $\Lambda$ and choosing the appropriate
integration constants~\cite{Thesis,YM-2-loop-A}, we arrive at the
relationship between the integrated \ERG\ kernels---\aka the
effective propagators---and the two-point,
tree level vertices shown in equation~\eq{eq:EPReln}. Note
that we have attached the effective propagator, which only
ever appears as an internal line, to an arbitrary structure
(this attachment ensures that the prescription for evaluating the group
theory factors, and not just the algebra, matches between the two sides
of the equation).
	\be
		\ensuremath{\begin{array}{c}\input{pstex/EffPropReln.pstex_t} \end{array}}
		\equiv \ensuremath{\begin{array}{c}\input{pstex/K-Delta.pstex_t} \end{array}} - \ensuremath{\begin{array}{c}\input{pstex/FullGaugeRemainder.pstex_t} \end{array}} 
		\equiv \ensuremath{\begin{array}{c}\input{pstex/K-Delta.pstex_t} \end{array}} - \ensuremath{\begin{array}{c}\input{pstex/DecomposedGR.pstex_t} \end{array}}
	\label{eq:EPReln}
	\ee

We have encountered $\GRk$ already, in section~\ref{sec:WIDs}. 
Of $\GRkpr$, all we need know for our purposes
is that it carries an index, $M$,
which is a four-index in the $A^1$, $A^2$ sectors, and a five index
in the $F$, $\bar{F}$ sectors. In the $C$ sector, $\GRkpr$ is null~\cite{YM-1-loop,Thesis,YM-2-loop-A}.
The structure $\GRkpr \!\! \GRk$
is a `gauge remainder'~\cite{YM-1-loop}. 
The individual components of
$\GRkpr \!\! \GRk$ will often be loosely
referred to as gauge remainders; where it is necessary to
 unambiguously refer to the composite structure, we will use
the terminology `full gauge remainder'.

It is important to note that we have defined the diagrammatics
in equation~\eq{eq:EPReln} such that 
there are no $1/N$
corrections where the effective propagator attaches
to the two-point, tree level vertex. We do this because, when the composite
object on the \lhs\ of equation~\eq{eq:EPReln} appears
in actual calculations, it always occurs inside some larger diagram.
It is straightforward to show that, in this case, the aforementioned attachment 
corrections always vanish~\cite{Thesis}. 

\subsubsection{Further Diagrammatic Identities}

The following diagrammatic identities all follow
from the ones stated already. However, since they
are so heavily used in perturbative calculations
we give them in their own right, particularly
as not all of them are immediately obvious.

The first of these, though, is trivial, 
following directly from~\eq{eq:D-ID-GR-TP-A}:
\be
	\ensuremath{\begin{array}{c}\input{pstex/GR-TLTP.pstex_t} \end{array}} = 0.
\label{eq:GR-TLTP}
\ee

From the effective propagator relation and~\eq{eq:GR-TLTP},
two further diagrammatic identities follow.
First, consider attaching
an effective propagator to the right-hand field in~\eq{eq:GR-TLTP}
and applying
the effective propagator before $\GRk$ has acted. Diagrammatically,
this gives
\[
	\ensuremath{\begin{array}{c}\input{pstex/GR-TLTP-EP.pstex_t} \end{array}} = 0 = \ensuremath{\begin{array}{c}\input{pstex/k.pstex_t} \end{array}} - \ensuremath{\begin{array}{c}\input{pstex/kkprk.pstex_t} \end{array}},
\]
which implies the following diagrammatic identity:
\be
	\ensuremath{\begin{array}{c}\input{pstex/GR-relation.pstex_t} \end{array}} = 1.
\label{eq:GR-relation}
\ee

The effective propagator relation, together
with~\eq{eq:GR-relation}, implies that
\[
	\ensuremath{\begin{array}{c}\input{pstex/TLTP-EP-GR.pstex_t} \end{array}} = \ensuremath{\begin{array}{c}\input{pstex/kpr.pstex_t} \end{array}} - \ensuremath{\begin{array}{c}\input{pstex/kprkkpr.pstex_t} \end{array}} = 0.
\]
In other words, the (non-zero) structure $\ensuremath{\begin{array}{c}\input{pstex/EP-GR.pstex_t} \end{array}}$ kills
a two-point, tree level vertex. But, by~\eq{eq:GR-TLTP}, 
this suggests that the structure $\ensuremath{\begin{array}{c}\input{pstex/EP-GR.pstex_t} \end{array}}$
must be equal, up to some factor, to $\lhd$. Indeed,
\be
	\ensuremath{\begin{array}{c}\input{pstex/EP-GRpr.pstex_t} \end{array}} \equiv \ensuremath{\begin{array}{c}\input{pstex/GR-PEP.pstex_t} \end{array}},
\label{eq:PseudoEP}
\ee
where the dot-dash line represents the pseudo effective propagators 
of~\cite{Thesis,YM-2-loop-A}.

\section{Universality}

To compute a universal quantity, we must feed in a
physical input, which is done via the renormalization
condition~\cite{GI-ERG-I,YM-1-loop}:
\be
	S[\SF = A^1, \SH = \bar{\SH}] =\frac{1}{2g^2} \, \tr \!\! \Int{x} \left(F_{\mu\nu}^1 \right)^2 + \cdots,
\label{eq:RenormCondition}
\ee
where the ellipsis denotes higher dimension operators and ignored vacuum energy and $\bar{\SH}$ is
the location of the minimum of the Higgs potential. 
This forces
\[
	S^{\ 1 \, 1}_{0 \mu \nu} (p) = 2 \Box_{\mu \nu}(p) + \Op{4},
\]
where the 1s are shorthand for $A^1$s and the $\Op{4}$ contributions to
the vertex are non-universal. We can arrange calculations
of universal $\beta$ function coefficients and, presumably, of
all universal quantities such that the answer is
manifestly 
controlled by the renormalization condition~\cite{Thesis,YM-2-loop-A,YM-2-loop-B,giqed}.

When computing $\beta_2$, we must remove all contributions
arising from the running of $\alpha$ in order to obtain
agreement with the standard, universal answer; this is done by
tuning $\alpha \rightarrow 0$ at the end of the calculation~\cite{Thesis,YM-2-loop-A,YM-2-loop-B}.\footnote{
Of course, disagreement with the standard value of $\beta_2$ is not
necessarily a signature of a sick formalism, since $\beta_2$ is
not physically observable and is only expected to agree between
two differing schemes under certain conditions.}  Note that
the $\gamma_i$ of~\eq{eq:gamma} are determined by the
renormalization condition for the unphysical coupling, $g_2$:
\[
	S[\SF = A^2, \SH = \bar{\SH}] = -\frac{1}{2\alpha g^2} \, \tr \!\! \Int{x} \left(F_{\mu\nu}^2 \right)^2 + \cdots.
\]

\section{Illustration}	\label{sec:Illustration}

To illustrate the diagrammatic techniques in action,
we will perform the initial stages of a computation
of $\beta_2$. To start, we specialize the weak coupling flow 
equations~\eq{eq:WeakFlow} to $n=2$, take $\{f\} = A^1_\mu A^1_\nu$
and work at $\Op{2}$. 
The renormalization condition~\eq{eq:RenormCondition}
implies that $S_{\geq 1 \mu \nu}^{\ \ \ 1 \, 1}(p) \sim \Op{4}$ and
so we are left with an algebraic equation for $\beta_2$,
as shown in figure~\ref{fig:beta2:L0}. The Lorentz indices
of the external fields are suppressed.
\bcf[h]
	\[
	-4\beta_2 \Box_{\mu \nu}(p) + \Op{4} = \frac{1}{2}
	\dec{
		\begin{array}{ccccc}
			\PD{L0-Sig_2}{fig:L0-Sig_2-Expand}	&	& \LD{WBT-Sig_1}	&						& \LD{L0-a0}
		\\[1ex]
			\ensuremath{\begin{array}{c}\input{pstex/Padlock-Sigma_1.pstex_t} \end{array}} 				& +	& \ensuremath{\begin{array}{c}\input{pstex/WBT-Sigma_1.pstex_t} \end{array}}	& \ds -	\sum_{r=0}^2 	& \ensuremath{\begin{array}{c}\input{pstex/Dumbbell-2_rb-rb.pstex_t} \end{array}}
		\end{array}
	}{11}
	\]
\caption{Diagrammatic equation for $\beta_2$.}
\label{fig:beta2:L0}
\ecf

Diagrams are labelled in boldface. If a diagram is cancelled, then
its reference number is enclosed in curly braces, together with
the reference number of the diagram against which it cancels.
If the reference number of a diagram
is followed by an arrow, the arrow points to the figure where
the diagram is processed. 
Since we are not performing the
complete diagrammatics for $\beta_2$, not all diagrams are
labelled and, of those that are, not all are processed or
cancelled.

When explicitly decorating with the external fields, we note that they
are identical, by Bose symmetry. Thus, for terms in which
the two fields decorate separate structures, we can simply
draw a single diagram but pick up a factor of two.

For each diagram generated by the flow, our strategy is as follows.
First, we isolate the component for which all vertices are Wilsonian
effective action vertices and for which the kernel is undecorated, should
it exist. To facilitate this separation, we introduce the 
symbol $\odot$ to indicate an undecorated kernel and $\circ$ 
such that, when operating on a kernel, $\bullet = \odot + \circ$.
In certain circumstance, we will be able to trade symbols; for
example, if a kernel carries $\circ$ but possesses an explicit
decoration, then we can replace $\circ$ by $\bullet$. 
The manipulable component of diagram~\ref{L0-Sig_2} is isolated
in figure~\ref{fig:L0-Sig_2-Expand}.
\bcf[h]
	\[
		\frac{1}{2}\dec{\ensuremath{\begin{array}{c}\input{pstex/Padlock-Sigma_1.pstex_t} \end{array}}}{11} = 
		\frac{1}{2}
		\dec{
			\begin{array}{ccccc}
				\PD{L0-S1-DEP}{fig:L0-LdL}	&	& \CD{L0:1-DW}{L2:1-DW}	&	& \CD{L0-hS1-W}{L1-hS1-W}
			\\[1ex]
				\ensuremath{\begin{array}{c}\input{pstex/Vertex-1-DEP.pstex_t} \end{array}} 			& +	& \ensuremath{\begin{array}{c}\input{pstex/Vertex-1-DW.pstex_t} \end{array}}		& -	& 2\ensuremath{\begin{array}{c}\input{pstex/Padlock-hS_1.pstex_t} \end{array}}
			\end{array}
		}{11}
	\]
\caption{Isolating the manipulable component for diagram~\ref{L0-Sig_2}.}
\label{fig:L0-Sig_2-Expand}
\ecf

Notice that this diagrammatic step has been performed without
explicitly decorating terms with the external fields,
thereby reducing the number of terms we have to deal with. This
is the first refinement of the
computational methodology employed in~\cite{YM-2-loop-A}. 

Next, there follows a two-step process.
First, we convert diagram~\ref{L0-S1-DEP} into a 
$\Lam$-derivative term, by moving the $\Lam$-derivative
off the effective propagator. This is shown in the first line
of figure~\ref{fig:L0-LdL}. On the second line, we promote the effective
propagator to an implicit decoration;
this promotion is the second refinement of the diagrammatics.
\bcf[h]
	\beas
		\frac{1}{2}\dec{\ensuremath{\begin{array}{c}\input{pstex/Vertex-1-DEP.pstex_t} \end{array}}}{11} & = &
		\frac{1}{2}
		\dec{
			\dec{\ensuremath{\begin{array}{c}\input{pstex/Vertex-1-EP.pstex_t} \end{array}}}{\bullet} -	\ensuremath{\begin{array}{c}\input{pstex/Vertex-LdL-EP.pstex_t} \end{array}}
		}{11}
	\\[1ex]
		& \equiv &
		\frac{1}{2}
		\dec{
			\begin{array}{ccc}
				\LD{L0:a1-LdL}					&	& \PD{L0:a1-LdL-Corr}{fig:L1:a1-Pro}
			\\[1ex]
				\dec{\ensuremath{\begin{array}{c}\input{pstex/Vertex-1.pstex_t} \end{array}}}{\bullet}	& -	& \ensuremath{\begin{array}{c}\input{pstex/Vertex-1-LdL.pstex_t} \end{array}}
			\end{array}
		}{11\Delta}
	\eeas
\caption{Converting diagram~\ref{L0-S1-DEP} into a $\Lam$-derivative term.}
\label{fig:L0-LdL}
\ecf

Notice that we have replaced $\odot$ by $\bullet$, since
we take these symbols to mean the same thing, \ie just
$-\flowConstAl$, when operating on a vertex or effective
propagator.

Diagram~\ref{L0:a1-LdL} is a $\Lam$-derivative term. 
The vertex is enclosed in square brackets which
tells us that $-\flowConstAl$ is taken to act \emph{after explicit decoration}.
However, in diagram~\ref{L0:a1-LdL-Corr}, it is just the vertex which is
struck by $-\flowConstAl$; this term is processed using the flow
equation, as shown in figure~\ref{fig:L1:a1-Pro}.
\bcf[h]
	\[
	\begin{array}{c}
		\ds
		-\frac{1}{2}\dec{\ensuremath{\begin{array}{c}\input{pstex/Vertex-1-LdL.pstex_t} \end{array}}}{11\Delta} =
	\\[4ex]
		\ds 
		-\frac{1}{2}
		\dec{
			\begin{array}{c}
				\ds
				-2\left[\beta_1 - \gamma_1 \pder{}{\alpha}\right] \ensuremath{\begin{array}{c}\input{pstex/Vertex-0.pstex_t} \end{array}}
			\\[4ex]
				\ds
				+\frac{1}{2}
				\left[
					\begin{array}{cccccc}
										& \PD{L1-Db-1b_r-rb}{fig:L1:a1:a0:I}	&	& \LD{L1:Sig_0-W}		&
					\\[1ex]
						\ds\sum_{r=0}^1	& \ensuremath{\begin{array}{c}\input{pstex/Dumbbell-1b_r-rb.pstex_t} \end{array}}					& -	& \ensuremath{\begin{array}{c}\input{pstex/Padlock-Sigma_0.pstex_t} \end{array}}	& -	& \ensuremath{\begin{array}{c}\input{pstex/WBT-Sigma_0.pstex_t} \end{array}}
					\end{array}
				\right]
			\end{array}
		}{11\Delta}
	\end{array}
	\]
\caption{Result of processing diagram~\ref{L0:a1-LdL-Corr} with the 
one-loop flow equation.}
\label{fig:L1:a1-Pro}
\ecf

There now follows a crucial step in the diagrammatic procedure:
we recognize that diagram~\ref{L1-Db-1b_r-rb} possesses
two-point, tree level vertices that can be attached to either
the external fields or the effective propagator. In the former case,
this generates a structure which is manifestly $\Op{2}$, allowing
us to Taylor expand at least some of the diagram's other structures in 
$p$. In
the latter case, we can apply the effective propagator relation. 

To facilitate the separation of the two-point, tree level
vertices, we define the reduced, $n$-loop vertex thus:
\[
	v^{R}_n =
	\left\{
		\begin{array}{cc}
			v_n & \mathrm{n>0} 
		\\
			v_0 - v^{\ XX}_{0 RS}(k) & \mathrm{n=0}
		\end{array}
		\right.,
\]
where we have suppressed all arguments of the generic 
vertex, $v_n$, and its reduction.
By definition, the reduced vertex does not contain a two-point, 
tree level component.
A superscript number in a vertex argument denotes the total number of fields
which must decorate the given vertex \eg $0^2$ is the
argument of a two-point, tree level vertex. Using this
notation, 
we isolate the two-point, tree
level vertices of diagram~\ref{L1-Db-1b_r-rb} in figure~\ref{fig:L1:a1:a0:I}.
\bcf[h]
	\[
	-\frac{1}{4} \sum_{r=0}^1 \dec{\ensuremath{\begin{array}{c}\input{pstex/Dumbbell-1b_r-rb.pstex_t} \end{array}}}{11\Delta}=
	-\frac{1}{4}
	\dec{
		\begin{array}{cccc}
							& \PD{L1:a1:a0-bR}{fig:L1:a1:a0-Separate}	&	& \PD{L1:a1:a0-TLTPb}{fig:a1:a0:TLTP-D}
		\\[1ex]
			\ds
			\sum_{r=0}^1 	& \ensuremath{\begin{array}{c}\input{pstex/Dumbbell-1b_rR-rbR.pstex_t} \end{array}}					& +2& \ensuremath{\begin{array}{c}\input{pstex/Dumbbell-1bR-TLTPb.pstex_t} \end{array}}
		\end{array}
	}{11\Delta}
	\]
\caption{Isolation of the two-point, tree level vertices of diagram~\ref{L1-Db-1b_r-rb}.}
\label{fig:L1:a1:a0:I}
\ecf

Diagram~\ref{L1:a1:a0-TLTPb} simplifies. First, we note that $\bar{1}^R = \bar{1}$, by
definition. Secondly, we note that
\be
	a_0[\bar{S}_n,\bar{S}_0^2] \equiv a_0[S_n,S_0^2] - a_0[S_n,\hS_0^2] - a_0[\hat{S}_n,S_0^2] = - a_0[\hat{S}_n,S_0^2],
\label{eq:Barred-TLTP}
\ee 
where the last step follows from the equality of the Wilsonian
effective action and seed action two-point, tree level vertices.
Performing this simplification, we decorate the two-point,
tree level vertex, to give the diagrams of figure~\ref{fig:a1:a0:TLTP-D}.
\bcf[h]
	\[
	-\frac{1}{2} \dec{\ensuremath{\begin{array}{c}\input{pstex/Dumbbell-1bR-TLTPb.pstex_t} \end{array}}}{11\Delta} =
	\dec{
		\begin{array}{ccc}
			\PD{L1:a1:a0-TLTP-WBT}{fig:a1:a0:EPReln}&	& \PD{L1:a1:a0-TLTP-EP}{fig:a1:a0:EPReln}
		\\[1ex]
			\ensuremath{\begin{array}{c}\input{pstex/Dumbbell-h1-WBT-TLTP-EP.pstex_t} \end{array}}			& +	& \ensuremath{\begin{array}{c}\input{pstex/Dumbbell-W-TLTP-EP.pstex_t} \end{array}}
		\end{array}
	}{11}
	+\frac{1}{2}
	\dec{
		\PDi{Dumbbell-h1-W-TLTP-E}{L1:a1:a0-TLTP-E}{fig:Op2}
	}{1\Delta}
	\]
\caption{Decorating the two-point, tree level vertex of diagram~\ref{L1:a1:a0-TLTPb}.}
\label{fig:a1:a0:TLTP-D}
\ecf

The relative factor of two in diagrams~\ref{L1:a1:a0-TLTP-WBT}
and~\ref{L1:a1:a0-TLTP-EP} comes from having been able to 
attach the effective propagator either way round. Explicit
decoration with an external $A^1$ is denoted by a wiggly
line, as exemplified in diagram~\ref{L1:a1:a0-TLTP-E}.
Were we to decorate this diagram with the 
remaining external field, we would
pick up a factor of two, since the external
fields would appear on different structures.

Diagrams~\ref{L1:a1:a0-TLTP-WBT}
and~\ref{L1:a1:a0-TLTP-EP} can be processed
using the effective propagator relation~\eq{eq:EPReln},
to give the terms of figure~\ref{fig:a1:a0:EPReln}.
\bcf[h]
	\beas
		\dec{\ensuremath{\begin{array}{c}\input{pstex/Dumbbell-h1-WBT-TLTP-EP.pstex_t} \end{array}}}{11}	& = &
		\dec{
			\begin{array}{ccc}
				\LD{L1-hS1-WBT}	&	& \PD{L1-hS1-WBT-GR}{fig:L1:GRs}
			\\[1ex]
				\ensuremath{\begin{array}{c}\input{pstex/WBT-hS_1.pstex_t} \end{array}}	& -	& \ensuremath{\begin{array}{c}\input{pstex/WBT-hS_1-GR.pstex_t} \end{array}}
			\end{array}
		}{11}
	\\[1ex]
		\dec{\ensuremath{\begin{array}{c}\input{pstex/Dumbbell-W-TLTP-EP.pstex_t} \end{array}}}{11}		& = &
		\dec{
			\begin{array}{ccc}
				\CD{L1-hS1-W}{L0-hS1-W}		&	& \PD{L1-hS1-W-GR}{fig:L1:GRs}
			\\[1ex]
				\ensuremath{\begin{array}{c}\input{pstex/Padlock-hS_1.pstex_t} \end{array}}			& -	& \ensuremath{\begin{array}{c}\input{pstex/Padlock-hS_1-GR.pstex_t} \end{array}}
			\end{array}
		}{11}
	\eeas
\caption{Processing diagrams~\ref{L1:a1:a0-TLTP-WBT}
and~\ref{L1:a1:a0-TLTP-EP} with the effective 
propagator relation.}
\label{fig:a1:a0:EPReln}
\ecf

The structure of the calculation now begins to reveal itself,
as we find two cancellations, which completely remove all
seed action contributions generated by the action of $a_1$
in figure~\ref{fig:beta2:L0}.

\begin{cancel}
Diagram~\ref{L1-hS1-WBT} exactly cancels the seed action contribution to diagram~\ref{WBT-Sig_1}.

\end{cancel}

\Cancel{L1-hS1-W}{L0-hS1-W}

The benefits of the refinements to the diagrammatic procedure
are now becoming clear: had we explicitly decorated \eg 
diagrams~\ref{L0-hS1-W} and~\ref{L1-hS1-W} with the external fields,
then the single cancellation~\ref{cancel:L1-hS1-W} would be replaced
by three separate cancellations; compared to the methodology 
of~\cite{YM-2-loop-A}, we have succeeded in cancelling these terms
in parallel. As the diagrammatic procedure is iterated, generating
diagrams with increasing numbers of vertices, the number of
terms cancelled in parallel grows rapidly, as we will see.

The gauge remainders of diagrams~\ref{L1-hS1-WBT-GR}
and~\ref{L1-hS1-W-GR} can be processed using the
techniques of section~\ref{sec:WIDs}, to yield
the terms of figure~\ref{fig:L1:GRs}. 
The notation has been adapted since the active field, being
an internal field, sits not only on the vertex struck by the
gauge remainder but also at the end of a kernel.
This kernel attaches to $\GRkpr$ and so,
rather than using the half-arrow notation of section~\ref{sec:WIDs},
we use the $\GRkpr$ to naturally indicate the momentum flow.
\bcf[h]
	\beas
		-\dec{\ensuremath{\begin{array}{c}\input{pstex/WBT-hS_1-GR.pstex_t} \end{array}}}{11}		& =	&
		2
		\dec{
			\begin{array}{ccc}
				\CD{GR-WBB-A}{GR-WBB-B}	& 	& \LD{L1-h1-W-hook}
			\\
				\ensuremath{\begin{array}{c}\input{pstex/Vertex-hS_1-WBB.pstex_t} \end{array}}	& -	& \ensuremath{\begin{array}{c}\input{pstex/Vertex-hS_1-WBT.pstex_t} \end{array}}
			\end{array}
		}{11}
	\\[1ex]
		-\dec{\ensuremath{\begin{array}{c}\input{pstex/Padlock-hS_1-GR.pstex_t} \end{array}}}{11}	& =	&
		-2
		\dec{
			\CDi{Vertex-hS_1-WBB}{GR-WBB-B}{GR-WBB-A}
		}{11}
		+4
		\dec{
			\CDi{Vertex-hS_1-WBE}{L2:h1-WBE}{L1:h1-WBE}
		}{1}
	\eeas
\caption{Result of processing the gauge remainders of diagrams~\ref{L1-hS1-WBT-GR}
and~\ref{L1-hS1-W-GR}.}
\label{fig:L1:GRs}
\ecf
Following section~\ref{sec:CC},
we have used charge conjugation to collect terms \ie the push forward
and pull back
onto a given field have been combined. In diagram~\ref{L2:h1-WBE},
an additional factor of two arises because the gauge remainder can strike
either of the external fields. Note that decoration with the
remaining external field now just yields a factor of unity.

\Cancel{GR-WBB-B}{GR-WBB-A}

Our next task is to examine diagram~\ref{L1:a1:a0-TLTP-E}.
Since we are working at $\Op{2}$, and this diagram
possesses a structure which is manifestly $\Op{2}$,
we expect to be able to directly Taylor expand the rest of the diagram
to zeroth order in $p$. 
In the case of this particular diagram, 
this na\"ive expectation is correct. More generally, individual
diagrams with an $\Op{2}$ stub may not be Taylor expandable in 
$p$; rather, only sums of diagrams can be Taylor 
expanded~\cite{Thesis,YM-2-loop-B}, since this
procedure can generate IR divergences in individual terms. For reasons that will
become apparent, it is actually only worth performing a Taylor
expansion in the case that the kernel is decorated (before
the Taylor expansion is performed). Diagram~\ref{L1:a1:a0-TLTP-E}
is processed in figure~\ref{fig:Op2}.
\bcf[h]
	\[
		\frac{1}{2}\dec{\ensuremath{\begin{array}{c}\input{pstex/Dumbbell-h1-W-TLTP-E.pstex_t} \end{array}}}{1\Delta}
		\rightarrow
		\frac{1}{2}
		\dec{
			\LDi{Dumbbell-h1-DEP-TLTP-E}{L1:a1:a0-TLTP-DEP}
		}{1\Delta}
		\begin{array}{cccc}
				& \LD{L1:a1:a0-h1-dW-TLTP-E}	&	& \LD{L1:a1:a0-dh1-DW-TLTP-E}
		\\[1ex]
			+2	& \ensuremath{\begin{array}{c}\input{pstex/Dumbbell-h1-dW-TLTP-E.pstex_t} \end{array}}	&+2	& \ensuremath{\begin{array}{c}\input{pstex/Dumbbell-dh1-DW-TLTP-E.pstex_t} \end{array}}
		\end{array}
	\]
\caption{Processing diagram~\ref{L1:a1:a0-TLTP-E} using the 
techniques of section~\ref{sec:Taylor}.}
\label{fig:Op2}
\ecf

There are two things to note. First, in diagrams~\ref{L1:a1:a0-h1-dW-TLTP-E} 
and~\ref{L1:a1:a0-dh1-DW-TLTP-E}
the top end of the kernel  carries zero momentum 
(at $\Op{2}$) and the bottom end carries
the same momentum as the decorative field (not to
be confused with the derivative symbol). We have thus combined terms
as discussed under equation~\eq{eq:Momderivs}. Secondly, we have not
included any diagrams in which the kernel is decorated by
the external field and either one loop or no loops. As
we know from section~\ref{sec:CC}, all
such diagrams vanish by charge conjugation invariance,
upon recognizing that the internal field attached to
the two-point, tree level vertex must be an $A^1$.

The strategy is now very simple: we iterate the
whole procedure. For each of diagrams~\ref{L0-a0}, 
\ref{L1:Sig_0-W} and~\ref{L1:a1:a0-bR}, we can separate
off a manipulable component, which we then convert into
a $\Lambda$-derivative term. Amongst the terms generated
will be a dumbbell structure, possessing at least one
two-point, tree level vertex. Decorating the two-point,
tree level vertices, we can either use the effective
propagator relation or we can perform
manipulations at $\Op{2}$.
There is one subtlety concerning the terms generated when
we perform the conversion into $\Lam$-derivative terms,
which can be illustrated by considering diagram~\ref{L1:a1:a0-bR};
we isolate the manipulable component in figure~\ref{fig:L1:a1:a0-Separate}.
In figure~\ref{fig:L1:a1:a0-P}, we convert diagram~\ref{L1:a1:a0-M}
into a $\Lam$-derivative term plus corrections.
\bcf[h]
	\[
	\begin{array}{c}
	\vspace{2ex}
		\ds
		-\frac{1}{4} \dec{\sum_{r=0}^1 \ensuremath{\begin{array}{c}\input{pstex/Dumbbell-1b_rR-rbR.pstex_t} \end{array}}}{11\Delta} = 
	\\
		\ds
		-\frac{1}{2}
		\dec{
			\begin{array}{ccccccc}
				\PD{L1:a1:a0-M}{fig:L1:a1:a0-P}	&	& \LD{L1:a1:a0-DW}		&	& \LD{L1:a1:a0-h1}		&	& \CD{L1:a1:a0:h0R}{L2-1-h0R}
			\\[1ex]
				\ensuremath{\begin{array}{c}\input{pstex/Dumbbell-1R-DEP-0R.pstex_t} \end{array}} 		& +	& \ensuremath{\begin{array}{c}\input{pstex/Dumbbell-1R-DW-0R.pstex_t} \end{array}}& -	& \ensuremath{\begin{array}{c}\input{pstex/Dumbbell-h1-0R.pstex_t} \end{array}} 	& -	& \ensuremath{\begin{array}{c}\input{pstex/Dumbbell-1-h0R.pstex_t} \end{array}}
			\end{array}
		}{11\Delta}
	\end{array}
	\]
\caption{A re-expression of diagram~\ref{L1:a1:a0-bR}.}
\label{fig:L1:a1:a0-Separate}
\ecf

\bcf[h]
	\[
	\begin{array}{c}
		\ds
		-\frac{1}{2}\dec{\ensuremath{\begin{array}{c}\input{pstex/Dumbbell-1R-DEP-0R.pstex_t} \end{array}}}{11\Delta} =
		+\frac{1}{4}
		\dec{
			\begin{array}{ccc}
				\LD{L1-1-DEP-0R}					&	& \LD{L1-1-0R-DEP}
			\\[1ex]
				\scd[4]{Vertex-1-DEP}{Vertex-0R}	& +	& \scd[4]{Vertex-1}{Vertex-0R-DEP}
			\end{array}
		}{11\Delta}
	\\[6ex]
		\ds
		-\frac{1}{8}
		\dec{
			\dec{\LO{\scd[4]{Vertex-1}{Vertex-0R}}{L1:a1:a0-LdL}}{\bullet}
			- 
			\left[
				\begin{array}{ccc}
					\LD{L1-1-LdL-0R}							&	& \PD{L1-1-0R-LdL}{fig:L1:a1:a0:a0-0R-flow}
				\\[1ex]
					\scd[4]{Vertex-1-LdL}{Vertex-0R}			& +	& \scd[4]{Vertex-1}{Vertex-0R-LdL}
				\end{array}
			\right]
		}{11\Delta^2}
	\end{array}
	\]
\caption{Converting diagram~\ref{L1:a1:a0-M} into a $\Lam$-derivative term plus corrections.}
\label{fig:L1:a1:a0-P}
\ecf

To proceed, we process both diagrams~\ref{L1-1-LdL-0R}
and~\ref{L1-1-0R-LdL}, using the weak coupling flow equations. This
is straightforward in the former case and we will not do it explicitly.
In the latter case,
we must understand how to compute the flow of
a reduced, (tree level) vertex. The point is that a reduced vertex lacks a
two-point, tree level component, and so the flow of a reduced vertex must lack
the flow of a two-point, tree level vertex. From section~\ref{sec:EPReln}, we
know that the flow of a two-point, tree level vertex generates two two-point,
tree level vertices, joined together by an undecorated kernel. Hence,
the flow of a reduced tree level vertex must generate a dumbbell structure
for which either at least one vertex is reduced or for which the kernel is
decorated. Simplifying the barred notation, where possible,  according 
to~\eq{eq:Barred-TLTP}, we obtain the diagrams of figure~\ref{fig:L1:a1:a0:a0-0R-flow}.
\bcf[h]
	\[
		\frac{1}{8} \dec{\scd[4]{Vertex-1}{Vertex-0R-LdL}}{11\Delta^2} 
		=
		\frac{1}{16}
		\dec{
			\begin{array}{ccccc}
				\LD{L2:1-0bR^2}						&	& \PD{L2:1-0hR-02}{fig:L2:TLTP-Dec}		&	& \PD{L2:1-02-02}{fig:L2:TLTP-Dec}
			\\[1ex]
				\scd[2]{Vertex-1}{Dumbbell-0bR-0bR}	& -2& \scd[2]{Vertex-1}{Dumbbell-0hR-02}	& - & \scd[2]{Vertex-1}{Dumbbell-02-02}
			\end{array}
		}{11\Delta^2}
	\]
\caption{Processing diagrams~\ref{L1-1-LdL-0R} and~\ref{L1-1-0R-LdL} using the tree-level flow equations.}
\label{fig:L1:a1:a0:a0-0R-flow}
\ecf

The next step is the decoration of the two-point, tree level vertices of
diagrams~\ref{L2:1-0hR-02} and~\ref{L2:1-02-02}. Not
all of the resultant terms are drawn, but rather
the selection shown in figure~\ref{fig:L2:TLTP-Dec}.
\bcf[h]
	\[
	\begin{array}{c}
	\vspace{2ex}
		\ds
		-\frac{1}{16} \dec{2\scd[2]{Vertex-1}{Dumbbell-0hR-02} + \scd[2]{Vertex-1}{Dumbbell-02-02}}{11\Delta^2} =
	\\
		\ds
		-\frac{1}{2}
		\dec{
			\begin{array}{cccccccc}
				\CD{L2-1-h0R}{L1:a1:a0:h0R}	& 	& \LD{L2-0hR-W;1}					&	& \CD{L2:1-DW}{L0:1-DW}	&	& \LD{L2:1-WBT}
			\\[1ex]
				\ensuremath{\begin{array}{c}\input{pstex/Dumbbell-1-h0R.pstex_t} \end{array}}			& +	& \scd[2]{Vertex-1}{Vertex-OhR-W}	& +	& \ensuremath{\begin{array}{c}\input{pstex/Vertex-1-DW.pstex_t} \end{array}}		&+2	& \ensuremath{\begin{array}{c}\input{pstex/WBT-1.pstex_t} \end{array}}
			\end{array}
		}{11\Delta} + \cdots
	\end{array}
	\]
\caption{Partial decoration of diagrams~\ref{L2:1-0hR-02} and~\ref{L2:1-02-02}.}
\label{fig:L2:TLTP-Dec}
\ecf

As expected, we find cancellations.

\Cancel{L2-1-h0R}{L1:a1:a0:h0R}
\Cancel{L2:1-DW}{L0:1-DW}

Had we explicitly decorated diagrams~\ref{L2-1-h0R}
and~\ref{L1:a1:a0:h0R} with the external fields
and effective propagator, then the single 
cancellation~\ref{cancel:L2-1-h0R} would be
replaced with twenty-four separate cancellations!\footnote{
For this particular diagram, this number could be considerably
reduced by working at $\Op{2}$ 
and noting that neither one-point, tree level vertices 
nor one-point, Wilsonian effective action vertices (of any loop order) exist.
However, the effects of such considerations must be
worked out on a diagram-by-diagram level. Furthermore,
in the current approach, there is no need to apply such
constraints at this stage of the diagrammatics, since
the cancellations are blind to such details, anyway.
Ultimately, the constraint that Wilsonian effective
action one-point vertices do not exist need only
be used to simplify the final set of $\Lambda$-derivative
terms.
}
Given that the two-loop diagrammatics can
ultimately generate diagrams possessing four
vertices and five internal lines, it is
clear what a huge simplification the new techniques
offer over the old methodology.

Notice, though, that diagram~\ref{L2:1-WBT}
does not exactly cancel the Wilsonian effective
action component of diagram~\ref{WBT-Sig_1}:
to complete the cancellation, we must
process diagram~\ref{L0-a0}. Specifically,
we should focus on the $r=1$ term. A sequence
of terms derived from this diagram is shown in
figure~\ref{fig:Sequence}, culminating in 
the diagram we require to complete the cancellation
of the Wilsonian effective action component of
diagram~\ref{WBT-Sig_1}.
\bcf[h]
	\[
	\begin{array}{c}
	\vspace{2ex}
		\ds
		-\frac{1}{2}\sum_{r=0}^2 \dec{\ensuremath{\begin{array}{c}\input{pstex/Dumbbell-2_rb-rb.pstex_t} \end{array}}}{11} \rightarrow 
		-\frac{1}{2}\dec{\ensuremath{\begin{array}{c}\input{pstex/Dumbbell-1-DEP-1.pstex_t} \end{array}}}{11} + \cdots \rightarrow
		\frac{1}{2}\dec{\scd[4]{Vertex-1-LdL}{Vertex-1}}{11\Delta} + \cdots
	\\
		\ds
		\rightarrow
		-\frac{1}{4} \dec{\scd[4]{WBT-Sigma_0}{Vertex-1}}{11\Delta} + \cdots \rightarrow
		\frac{1}{4} \dec{\scd[4]{WBT-02}{Vertex-1}}{11\Delta} + \cdots \rightarrow
		\frac{1}{2} \dec{\ensuremath{\begin{array}{c}\input{pstex/WBT-1.pstex_t} \end{array}}}{11} +\cdots
	\end{array}
	\]
\caption{A sequence of terms spawned by diagram~\ref{L0-a0}.}
\label{fig:Sequence}
\ecf

Iterating the diagrammatic procedure until exhaustion,
we find that, up to gauge remainder terms and terms
that require manipulation at $\Op{2}$, the calculation
reduces to $\alpha$, $\beta$ and $\Lam$-derivative terms.
The only cancellations involved in this
procedure that we have not seen are those which
remove terms such as~\ref{L1-1-DEP-0R}, \ref{L1-1-0R-DEP} 
and~\ref{L2-0hR-W;1}.

The way in which these terms are cancelled is simple. Notice that 
bottom two structures of
the latter two diagrams combine to form a contribution 
to $\Sigma_0$; 
were we to  perform the complete diagrammatics, we 
would find the missing components.
The resulting diagram would then cancel against
a term spawned from the manipulation of diagram~\ref{L0-a0};
indeed, this term is included in the ellipsis
after the fourth term of figure~\ref{fig:Sequence}.

We conclude our illustration of the diagrammatic techniques
with some further examples of gauge remainders
and manipulations at $\Op{2}$. First,
consider the diagrams of figure~\ref{fig:Ex-GRs}.
\bcf[h]
	\[
		\frac{1}{2}
		\dec{
			\ensuremath{\begin{array}{c}\input{pstex/Dumbbell-h1-W-GR-0R.pstex_t} \end{array}}
		}{11\Delta} 
		=
		-\dec{
			\ensuremath{\begin{array}{c}\input{pstex/Vertex-h1-W-O-St.pstex_t} \end{array}}
		}{11\Delta}
		=
		-
		\dec{
			\begin{array}{ccc}
										&	& \PD{GR-Ex-A}{fig:Nested}
			\\[1ex]
				\ensuremath{\begin{array}{c}\input{pstex/Vertex-h1-W-OR-St.pstex_t} \end{array}}	& +	& \ensuremath{\begin{array}{c}\input{pstex/Vertex-h1-W-O2-St.pstex_t} \end{array}}
			\end{array}
		}{11\Delta}
	\]
\caption{Example of a gauge remainder generated from iterating the diagrammatic procedure.}
\label{fig:Ex-GRs}
\ecf

On the \lhs, we have a gauge remainder term. On the
\rhs, we have allowed the gauge remainder to
act but have not specified which field it strikes,
by employing the socket notation of~\cite{Thesis}.
This socket can be filled by any of the
decorations.
We have combined the push forward and pull back onto
the socket, using charge conjugation invariance, choosing
to represent the pair as a pull back (hence the factor of
minus two, compared to the parent).

Notice that a gauge remainder striking a reduced vertex
generates a full vertex. This is trivial to see: since
two-point, tree level vertices are killed by gauge remainders,
we can promote a reduced vertex struck by a gauge remainder 
to a full vertex. Given
that the action of the gauge remainder generates
a full two-point, tree level vertex, our strategy is as
before: we isolate any two-point, tree level contributions
and partially decorate them. Amongst the terms generated
by this latter procedure are those of figure~\ref{fig:Nested}.
\bcf[h]
	\[
	-\dec{\ensuremath{\begin{array}{c}\input{pstex/Vertex-h1-W-O2-St.pstex_t} \end{array}}}{11\Delta}
	= -4
	\dec{
		\begin{array}{ccc}
			\CD{L1:h1-WBE}{L2:h1-WBE}	&	& \LD{L2:h1-WBE-GR}
		\\[1ex]
			\ensuremath{\begin{array}{c}\input{pstex/Vertex-hS_1-WBE.pstex_t} \end{array}}		& -	& \ensuremath{\begin{array}{c}\input{pstex/Vertex-hS_1-WBE-GR.pstex_t} \end{array}}
		\end{array}
	}{1}
	\]
\caption{A selection of terms arising from the partial decoration
of diagram~\ref{GR-Ex-A}.}
\label{fig:Nested}
\ecf

\Cancel{L1:h1-WBE}{L2:h1-WBE}

Diagram~\ref{L2:h1-WBE-GR} is an example of a nested gauge remainder.
The action of the nested gauge remainder is exactly the same as
for any other gauge remainder. However, we recall 
from section~\ref{sec:CC}
that we cannot generally
use charge conjugation invariance to collect together diagrams
in which the nested gauge remainder has acted.

The next example will demonstrate how
diagrams involving processed gauge remainders can be 
converted into $\Lambda$-derivative terms. Consider
the diagrams in figure~\ref{fig:GR-LdL-pre}. Notice
that the first term is closely related to 
diagram~\ref{L1-h1-W-hook}.
\bcf[h]
	\[
		2
		\dec{
			\begin{array}{ccc}
				\PD{1-WBT}{fig:GR-LdL}	&	& \PD{1-EPX-TLTP-DEP-GR}{fig:GR-LdL}
			\\[1ex]
				\ensuremath{\begin{array}{c}\input{pstex/Vertex-1-WBT.pstex_t} \end{array}}		& -	& \ensuremath{\begin{array}{c}\input{pstex/Vertex-1-EPX-TLTP-DEP-GR.pstex_t} \end{array}}
			\end{array}
		}{11}
	\]
\caption{Two diagrams possessing processed gauge remainders
which can be converted into a $\Lam$-derivative term.}
\label{fig:GR-LdL-pre}
\ecf

Diagram~\ref{1-EPX-TLTP-DEP-GR} possesses a structure
we have not yet encountered: the two-point, tree level
vertex is attached exclusively to internal fields
but cannot be removed by the effective propagator
relation. Its top socket attaches to a differentiated
effective propagator, whereas the attachment of an
effective propagator to its bottom socket is interrupted
by a processed gauge remainder. However, we can make progress
by utilizing diagrammatic identities~\eq{eq:LdL-GRk-Pert-a}, 
\eq{eq:EPReln}--\eq{eq:PseudoEP}
and the tree level flow equation. We have:
\bea
	\ensuremath{\begin{array}{c}\input{pstex/TLTP-DEP-GR.pstex_t} \end{array}}	& =	& \dec{\ensuremath{\begin{array}{c}\input{pstex/TLTP-EP-GR-B.pstex_t} \end{array}}}{\bullet} - \ensuremath{\begin{array}{c}\input{pstex/TLTP-LdL-EP-GR.pstex_t} \end{array}} - \ensuremath{\begin{array}{c}\input{pstex/TLTP-EP-GR-LdL.pstex_t} \end{array}}	\nonumber
\\[1ex]
						& =	& \dec{\ensuremath{\begin{array}{c}\input{pstex/TLTP-GR-PEP.pstex_t} \end{array}}}{\bullet} - \ensuremath{\begin{array}{c}\input{pstex/TLTP-LdL-GR-PEP.pstex_t} \end{array}} - \ensuremath{\begin{array}{c}\input{pstex/GR-LdL.pstex_t} \end{array}} + \ensuremath{\begin{array}{c}\input{pstex/GR-GR-LdL.pstex_t} \end{array}} \nonumber
\\[1ex]
						& =	& - \ensuremath{\begin{array}{c}\input{pstex/GR-LdL.pstex_t} \end{array}} 
\label{eq:GR-LdL}
\eea

To go from the first line to the second, we have employed diagrammatic 
identity~\eq{eq:PseudoEP} and the effective
propagator relation. On the second line, the
first term vanishes courtesy of diagrammatic
identity~\eq{eq:GR-TLTP}; similarly, the second
term, if we employ~\eq{eq:LdL-GRk-Pert-a}. The final term on the second line
vanishes on account of diagrammatic identities~\eq{eq:GR-relation}
and~\eq{eq:LdL-GRk-Pert-a}:
\[
	\dec{ \GRk \!\! \GRkpr}{\bullet} = 0 = \stackrel{\bullet}{\GRk} \! \GRkpr + \GRk \! \! \stackrel{\bullet}{\GRkpr} = \GRk \!\!  \stackrel{\bullet}{\GRkpr}.
\]
It is thus apparent that we can re-write diagrams~\ref{1-WBT}
and~\ref{1-EPX-TLTP-DEP-GR}, as shown in figure~\ref{fig:GR-LdL}.
\bcf[h]
	\[
		2 \dec{\ensuremath{\begin{array}{c}\input{pstex/Vertex-1-WBT.pstex_t} \end{array}} -\ensuremath{\begin{array}{c}\input{pstex/Vertex-1-EPX-TLTP-DEP-GR.pstex_t} \end{array}}}{11} = 
		2 
		\dec{
		\begin{array}{ccc}
												&	& \LD{1-LdL-EB-hook}
		\\[1ex]
			\dec{\ensuremath{\begin{array}{c}\input{pstex/Vertex-1-EP-WBT.pstex_t} \end{array}}}{\bullet} & -	& \ensuremath{\begin{array}{c}\input{pstex/Vertex-1-LdL-EP-WBT.pstex_t} \end{array}}
		\end{array}
		}{11}
	\]
\caption{Conversion of diagrams~\ref{1-WBT}
and~\ref{1-EPX-TLTP-DEP-GR} into a $\Lam$-derivative
term.}
\label{fig:GR-LdL}
\ecf

Notice that, amongst the terms generated by processing
diagram~\ref{1-LdL-EB-hook}, is a term which
will cancel diagram~\ref{L1-h1-W-hook}.

By processing all the gauge remainders,
we can reduce the calculation to
a set of $\alpha$, $\beta$ and $\Lam$-derivative
terms, up to those diagrams which require manipulation
at $\Op{2}$ and a set of diagrams which do not
manifestly cancel. It is easy to find an example
of terms of the latter type. When diagram~\ref{L1-h1-W-hook}
is cancelled, we know that a gauge remainder term will be
left behind, as shown in figure~\ref{fig:surviving-GR}.
\bcf[h]
	\[
	-2
	\dec{
		\ensuremath{\begin{array}{c}\input{pstex/Vertex-1-W-GR-hook.pstex_t} \end{array}}
	}{11}
	\]
\caption{A gauge remainder term which cannot be processed.}
\label{fig:surviving-GR}
\ecf

The full gauge remainder is trapped, and cannot be processed.
The resolution to this problem is trivial, in this case:
by charge conjugation invariance, the (necessarily bosonic)
kernel has support only in the $C^1$ and $C^2$-sectors; but
these are precisely the sectors where gauge remainders are null.
$\cdeps{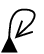} =0$ is a (trivial) example of a secondary diagrammatic
identity.
As we iterate the diagrammatic procedure, we find more
complicated examples. The first non-trivial case is~\cite{Thesis}
\[
	\ensuremath{\begin{array}{c}\input{pstex/Struc-AR1-A+B.pstex_t} \end{array}} - \cdeps{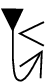} + \ensuremath{\begin{array}{c}\input{pstex/Struc-AR1-A+B-b.pstex_t} \end{array}} - \cdeps{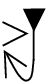} \equiv 0.
\]
This is one of a family of
secondary diagrammatic identities required
for the computation of $\beta_2$, which are given in~\cite{Thesis,YM-2-loop-A}.
The arbitrary loop generalization will be presented in~\cite{mgiuc}.

Finally, we will deal with an example involving
manipulations at $\Op{2}$, which in turn generate
gauge remainders. First, consider the fully decorated
diagram
on the \lhs\ of figure~\ref{fig:TE-Ex}, which we
manipulate at $\Op{2}$. To get to the
second line, we have recognized that
\[
	\ensuremath{\begin{array}{c}\input{pstex/TLTP-EP-d.pstex_t} \end{array}} = \ensuremath{\begin{array}{c}\input{pstex/TLTPd-EP.pstex_t} \end{array}} + \ensuremath{\begin{array}{c}\input{pstex/TLTP-EPd.pstex_t} \end{array}} = -\ensuremath{\begin{array}{c}\input{pstex/GRd.pstex_t} \end{array}},
\]
as a consequence of the effective propagator relation
(where, strictly, this is only true when the
structures involved are part of some complete diagram
\cf~\eq{eq:EPReln}).

\bcf[h]
	\beas
		-\ensuremath{\begin{array}{c}\input{pstex/TE-Ex-A.pstex_t} \end{array}}	&\rightarrow& - 2\ensuremath{\begin{array}{c}\input{pstex/TE-Ex-B.pstex_t} \end{array}}
	\\[1ex]
						& = 		& 2
		\left[
			\begin{array}{ccc}
				\LD{TE-Ex-C}	&	& \LD{TE-Ex-D}
			\\[1ex]
				\ensuremath{\begin{array}{c}\input{pstex/TE-Ex-C.pstex_t} \end{array}}	& +	& \ensuremath{\begin{array}{c}\input{pstex/TE-Ex-D.pstex_t} \end{array}}
			\end{array}
		\right]
	\eeas
\caption{Example of a diagram which can be manipulated at $\Op{2}$.}
\label{fig:TE-Ex}
\ecf

Diagrams~\ref{L1:a1:a0-h1-dW-TLTP-E}, \ref{L1:a1:a0-dh1-DW-TLTP-E}
and~\ref{TE-Ex-C} combine into a total derivative
\wrt\ the momentum flowing around the loop. However, the preregularization
necessary to properly define the $SU(N|N)$ regularization scheme~\cite{SU(N|N)}
ensures that such terms vanish. Clearly, this happens automatically
if we employ dimensional regularization as a preregularizer. However,
as we will explore more fully in~\cite{mgiuc}, it seems likely
that we can instead adopt a diagrammatic prescription whereby
sets of diagrams such as~\ref{L1:a1:a0-h1-dW-TLTP-E}, \ref{L1:a1:a0-dh1-DW-TLTP-E}
and~\ref{TE-Ex-C} can be discarded, purely from the standpoint
that they can be diagrammatically cast as a total momentum derivative.
This prescription would make sense in $D$ strictly equal to four.

With diagram~\ref{TE-Ex-D}, it looks as though we might be stuck,
since the gauge remainder is differentiated \wrt\ momentum. However,
trivially redrawing
\[
	\ensuremath{\begin{array}{c}\input{pstex/DiffGR.pstex_t} \end{array}} = \ensuremath{\begin{array}{c}\input{pstex/DiffGRk.pstex_t} \end{array}} + \ensuremath{\begin{array}{c}\input{pstex/DiffGRkpr.pstex_t} \end{array}},
\]
it is clear that progress can be made. In the case that the derivative hits
the $\GRkpr$, the $\GRk$ strikes the kernel  and can be processed
as usual.
In the case that the derivative
hits $\GRk$, we then use diagrammatic identity~\eq{eq:PseudoEP} to yield
a $\GRk$ striking the one-loop, seed action vertex. We will examine
the former case but, rather than dealing directly with diagram~\ref{TE-Ex-D},
will deal with the partner diagram (which we will
not explicitly generate), coming with opposite sign, 
in which the one-loop vertex is
a Wilsonian effective action vertex, rather than a seed action vertex.
We focus on the term in which the gauge remainder pulls back along
the kernel. Together with this diagram, we consider two diagrams which
can be manipulated at $\Op{2}$, as shown in figure~\ref{fig:TE-LdL-Ex}.
\bcf[h]
	\[
		2
		\dec{
			\begin{array}{ccccc}
				\PD{02-W-1-EP-dGR}{fig:TE-LdL-Ex-B}	&	& \PD{02EP-1-WE-GR}{fig:TE-LdL-Ex-B}&	& \PD{02-EP-1-EP-TLTh-W-GR}{fig:TE-LdL-Ex-B}
			\\[1ex]
				\ensuremath{\begin{array}{c}\input{pstex/TLTP-W-1-EP-dGR.pstex_t} \end{array}}				& -	& \ensuremath{\begin{array}{c}\input{pstex/TLTP-EP-1-WE-GR.pstex_t} \end{array}}				& -	& \ensuremath{\begin{array}{c}\input{pstex/TLTP-EP-1-EP-TLTh-W-GR.pstex_t} \end{array}}
			\end{array}
		}{}
	\]
\caption{A selection of three terms possessing an $\Op{2}$ stub.}
\label{fig:TE-LdL-Ex}
\ecf

Diagrammatically Taylor expanding the final two terms, we can
cast the three diagrams of figure~\ref{fig:TE-LdL-Ex} as a
$\Lam$-derivative term. This is most easily seen by noting that
\beas
	\ensuremath{\begin{array}{c}\input{pstex/GR-W-dTLTP-EP.pstex_t} \end{array}} 	& = & \ensuremath{\begin{array}{c}\input{pstex/GR-W-TLTP-d-EP.pstex_t} \end{array}} - \ensuremath{\begin{array}{c}\input{pstex/GR-dW-TLTP-EP.pstex_t} \end{array}} -\ensuremath{\begin{array}{c}\input{pstex/dGR-W-TLTP-EP.pstex_t} \end{array}} 
\\
						& = & - \ensuremath{\begin{array}{c}\input{pstex/LdLdGR-EP.pstex_t} \end{array}} -\ensuremath{\begin{array}{c}\input{pstex/GR-dW.pstex_t} \end{array}} + \ensuremath{\begin{array}{c}\input{pstex/GR-dW-GR.pstex_t} \end{array}} -\ensuremath{\begin{array}{c}\input{pstex/dGR-W.pstex_t} \end{array}} + \ensuremath{\begin{array}{c}\input{pstex/dGR-W-GR.pstex_t} \end{array}}.
\eeas
The first diagram on the second line follows in exactly the same way as~\eq{eq:GR-LdL}.
The other diagrams on the second line are simply obtained by using the effective
propagator relation. Applying the above relation to diagram~\ref{02-EP-1-EP-TLTh-W-GR} generates
five diagrams. The third and fifth die, since they involve a gauge remainder
striking a Wilsonian effective action, two-point vertex. The second diagram
cancels diagram~\ref{02EP-1-WE-GR}, after the latter diagram is manipulated at $\Op{2}$.
The remaining two diagrams combine with diagram~\ref{02-W-1-EP-dGR} to form a $\Lam$-derivative
term, as shown in figure~\ref{fig:TE-LdL-Ex-B}. Note that, at $\Op{2}$,
we can take the $\flowConstAl$ to strike the entire diagram \ie including
the two-point, tree level vertex decorated by the external field; this
follows because the $\Op{2}$ stub is independent of $\Lambda$.
\bcf[h]
	\[
	2 \dec{\ensuremath{\begin{array}{c}\input{pstex/TLTP-EP-1-EP-dGR.pstex_t} \end{array}}}{\bullet}
	-2 \LDi{TLTP-EP-1-LdL-EP-dGR}{02-EP-1-LdL-EP-dGR}
	\]
\caption{Rewriting diagrams~\ref{02-W-1-EP-dGR}--\ref{02-EP-1-EP-TLTh-W-GR} as
a $\Lam$-derivative term.}
\label{fig:TE-LdL-Ex-B}
\ecf

The final points can be made by considering manipulating
diagram~\ref{02-EP-1-LdL-EP-dGR}. We know that the flow
of the one-point vertex will generate, amongst other
terms, the usual dumbbell structure possessing a two-point,
tree level vertex. Applying the effective propagator
relation, the Kronecker-$\delta$ terms will cancel diagrams 
generated elsewhere, leaving
behind the trapped gauge remainders shown in figure~\ref{fig:TE-Trap}.
\bcf[h]
	\[
		-2
		\left[
			\begin{array}{ccc}
				\PD{02-EP-1-W-GR-dGR}{fig:TE-redraw}	&	& \PD{02-GR-W-1-EP-dGR}{fig:TE-redraw-B}
			\\[1ex]
				\ensuremath{\begin{array}{c}\input{pstex/TLTP-EP-1-W-GR-dGR.pstex_t} \end{array}}					& +	& \ensuremath{\begin{array}{c}\input{pstex/TLTP-GR-W-1-EP-dGR.pstex_t} \end{array}}
			\end{array}
		\right]
	\]
\caption{Trapped gauge remainders generated by processing diagram~\ref{02-EP-1-LdL-EP-dGR}.}
\label{fig:TE-Trap}
\ecf

Both diagrams~\ref{02-EP-1-W-GR-dGR} and~\ref{02-GR-W-1-EP-dGR} can be 
redrawn. In the former case, we first note from diagrammatic
identity~\eq{eq:GR-relation} that we can move the momentum derivative
from $\GRkpr$ to $\GRk$, at the expense of a minus sign. Now, since
it is true that, in all sectors for which the gauge remainder is not null,
\[
	\ensuremath{\begin{array}{c}\input{pstex/dGRk.pstex_t} \end{array}} = \delta_{\alpha\nu},
\]
we can redraw diagram~\ref{02-EP-1-W-GR-dGR} as shown
in figure~\ref{fig:TE-redraw}.
\bcf[h]
	\[
	\begin{array}{cccc}
			&											&			& \CD{02-EP-1-W-GR-GRE}{02-EP-1-W-GR-GRE-C}
	\\[1ex]
		-2 	&\ensuremath{\begin{array}{c}\input{pstex/TLTP-EP-1-W-GR-dGR.pstex_t} \end{array}} 					&\equiv 2	&\ensuremath{\begin{array}{c}\input{pstex/TLTP-EP-1-W-GR-GRE.pstex_t} \end{array}}
	\end{array}
	\]
\caption{Exact redrawing of diagram~\ref{02-EP-1-W-GR-dGR}.}
\label{fig:TE-redraw}
\ecf

Diagram~\ref{02-GR-W-1-EP-dGR} is redrawn,
as shown in figure~\ref{fig:TE-redraw-B}. 
\bcf[h]
	\[
	\begin{array}{cccccc}
			&							&	&								&	& \CD{02-GR-W-1-EP-dGR-rd}{02-GR-W-1-EP-dGR-rd-C}		
	\\[1ex]
		-2	& \ensuremath{\begin{array}{c}\input{pstex/TLTP-GR-W-1-EP-dGR.pstex_t} \end{array}}	&=-2& \ensuremath{\begin{array}{c}\input{pstex/TLTP-GR-W-1-EP-dGR-pre.pstex_t} \end{array}}	&=2	& \ensuremath{\begin{array}{c}\input{pstex/TLTP-GR-W-1-EP-dGR-rd.pstex_t} \end{array}}
	\end{array}
	\]
\caption{Exact redrawing of diagram~\ref{02-GR-W-1-EP-dGR}.}
\label{fig:TE-redraw-B}
\ecf

In the second diagram, if the left-most gauge remainder 
pushes forward, onto the internal field,
we regenerate the parent. If this gauge remainder pulls back onto the internal
field, the supertrace structure of the diagram is 
uniquely determined to be
$\str A^1_\mu \str A^1_\nu = 0$. If either 
gauge remainder strikes the external field,
we are left with a two-point, tree level vertex
struck by a gauge remainder, which vanishes by diagrammatic identity~\eq{eq:GR-TLTP}.
To go from the second diagram to the third, we allow the right-most gauge remainder 
to act; the only surviving contribution is the pull back onto the internal 
field, giving diagram~\ref{02-GR-W-1-EP-dGR-rd}.

Redrawing diagrams~\ref{02-EP-1-W-GR-dGR} and~\ref{02-GR-W-1-EP-dGR}
in this manner now allows us to understand how they are cancelled.
In figure~\ref{fig:TE-Cancel}, we show the parent diagram,
amongst the daughter diagrams of which, are the terms which
yield the cancellations we are looking for.
\bcf[h]
	\[
	\begin{array}{ccccccc}
			&										&					& \CD{02-EP-1-W-GR-GRE-C}{02-EP-1-W-GR-GRE}	&	& \CD{02-GR-W-1-EP-dGR-rd-C}{02-GR-W-1-EP-dGR-rd}	&
	\\
		-	& \ensuremath{\begin{array}{c}\input{pstex/Vertex-h1-W-GR-TLTh-EP-TlTh-EP.pstex_t} \end{array}}	& \rightarrow -2	& \ensuremath{\begin{array}{c}\input{pstex/TLTP-EP-1-W-GR-GRE.pstex_t} \end{array}}					&-2	& \ensuremath{\begin{array}{c}\input{pstex/TLTP-GR-W-1-EP-dGR-rd-B.pstex_t} \end{array}}						& +\cdots
	\end{array}
	\]
\caption{Generation of the diagrams to cancel~\ref{02-EP-1-W-GR-dGR} and~\ref{02-GR-W-1-EP-dGR}.}
\label{fig:TE-Cancel}
\ecf

\Cancel{02-EP-1-W-GR-GRE-C}{02-EP-1-W-GR-GRE}
\CancelCom{02-GR-W-1-EP-dGR-rd-C}{02-GR-W-1-EP-dGR-rd}{ upon recalling that
charge conjugation invariance allows us to reflect a diagram, picking up a sign
for each performed gauge remainder and each momentum derivative.}
These cancellations could have been performed directly against 
diagrams~\ref{02-EP-1-W-GR-dGR} and~\ref{02-GR-W-1-EP-dGR} by noting that
the redrawing of figures~\ref{fig:TE-redraw} and~\ref{fig:TE-redraw-B}
can be thought of as applications of (new)
secondary diagrammatic identities.
Cancellations~\ref{cancel:02-EP-1-W-GR-GRE-C}
and~\ref{cancel:02-GR-W-1-EP-dGR-rd-C} 
complete the illustration of the $\beta_2$
diagrammatics; the diagrammatic procedure
is summarized in the flow chart of 
figure~\ref{fig:flowchart}.
\bcf[t]
	\[
	\ensuremath{\begin{array}{c}\input{pstex/FlowChart.pstex_t} \end{array}}
	\]
\caption{The diagrammatic procedure.}
\label{fig:flowchart}
\ecf

Iterating the entire procedure until exhaustion,
we can reduce the calculation to a set of $\alpha$, $\beta$
and $\Lambda$-derivative terms and 
a set of `$\Op{4}$ terms'. Diagram~\ref{L1:a1:a0-TLTP-DEP}
is an example of a term of the latter type; it can be
easily demonstrated, using the Ward identities, 
that the sum of these terms
vanish at $\Op{2}$~\cite{Thesis,YM-2-loop-A}.
The $\Lambda$-derivative and $\beta$-terms
can be
simplified by utilizing the diagrammatic expression for
$\beta_1$~\cite{Thesis,YM-2-loop-B}, to yield an expression
for $\beta_2$ in terms of just $\alpha$ and $\Lambda$-derivative
terms. As mentioned already,
the $\alpha$-terms vanish in the limit that
$\alpha \to 0$. It is beyond the scope of this paper
to describe the extraction of the numerical coefficient
from the final set of terms; these techniques are fully
described in~\cite{Thesis,YM-2-loop-B} (see also~\cite{giqed}
for an example of their application in a simplified context).

\ack

I acknowledge financial support from PPARC.

\section*{References}

\input{./Bibliography/biblio.tex}
\end{document}

%% file: Preamble/preamble.tex
\newcounter{Diagrams}
\Alph{Diagrams}
\newcounter{cancellation}
\addtocounter{cancellation}{1}

\newtheorem{D}{}[Diagrams]
\newtheorem{DC}[D]{\{}
\newtheorem{cancel}{Cancellation}

\newcommand{\ERG}{ERG}

\newcommand{\eg}{e.g.\ }
\newcommand{\ie}{i.e.\ }
\newcommand{\cf}{cf.\ }
\newcommand{\etc}{etc.\ }
\newcommand{\rhs}{right-hand side}
\newcommand{\lhs}{left-hand side}
\newcommand{\wrt}{with respect to}
\newcommand{\aka}{a.k.a.\ }

\newcommand{\role}{r\^{o}le}

\newcommand{\be}{\begin{equation}}
\newcommand{\ee}{\end{equation}}
\newcommand{\bea}{\begin{eqnarray}}
\newcommand{\eea}{\end{eqnarray}}
\newcommand{\beas}{\begin{eqnarray*}}
\newcommand{\eeas}{\end{eqnarray*}}

\newcommand{\bcf}{\begin{center}\begin{figure}}
\newcommand{\ecf}{\end{figure}\end{center}}

\newcommand{\bct}{\begin{center}\begin{table}}
\newcommand{\ect}{\end{table}\end{center}}

\newcommand{\ds}{\displaystyle}

\newlength{\MinusLength}
\newcommand{\eq}[1]{(\ref{#1})}

\newcommand{\str}{\mathrm{str}\,}

\newcommand{\Int}[1]{\int \!\! d^D \! #1 \,}

\newcommand{\af}[1]{\bar{#1}}

\newcommand{\SF}{\mathcal{A}}
\newcommand{\SH}{\mathcal{C}}
\newcommand{\GRk}{\rhd}
\newcommand{\GRkpr}{>}

\newcommand{\Lam}{\Lambda}

\newcommand{\hS}{\hat{S}}

\newcommand{\bigdot}[1]{\stackrel{\bullet}{#1}}
\newcommand{\dd}{\bigdot{\Delta}}

\newcommand{\one}{\ensuremath{1\! \mathrm{l}}}

\newcommand{\pder}[2]{\ensuremath{\frac{\partial #1}{\partial #2}}}

\newcommand{\Op}[1]{\mathcal{O}(p^{#1})}

\newcommand{\flow}{\ensuremath{\Lambda \partial_\Lambda }}
\newcommand{\flowConstAl}{\ensuremath{\Lambda \partial_\Lambda|_\alpha }}
\newcommand{\dec}[3][0]{\ensuremath{\left[ #2 \hspace{#1in} \right]^{#3}}}

\newcommand{\DAD}{\scriptscriptstyle \odot}

\newcommand{\DiagDot}{\scriptstyle \bullet}
\newcommand{\DummyKernel}{\ensuremath{\stackrel{\bullet}{\mbox{\rule{1cm}{.2mm}}}}}

\newlength{\VertexWidth}
\newcommand{\Dal}{
	\ensuremath{
		\begin{array}{c}
			\settowidth{\VertexWidth}{\scriptsize $\alpha$}
			\setlength{\unitlength}{1.6\VertexWidth}
			\begin{picture}(1,1)(-0.5,-0.5)
				\put(0,0){\circle{1}}
				\put(-0.45,-0.25){$\alpha$}
			\end{picture}
		\end{array}
	}
}

\newlength{\LabLength}
\newlength{\ProcessRefLength}
\newlength{\ProcessLength}
\newlength{\CancelRefLength}
\newlength{\CancelLength}

\newcommand{\cdeps}[1]{\ensuremath{\begin{array}{c}\includegraphics{eps/#1.eps} \end{array}}}

\newcommand{\sco}[3][0]{
	\begin{array}{c}
		#2 \\[#1ex]
		#3
\end{array}
}

\newcommand{\scd}[3][0]{
	\sco[#1]{\ensuremath{\begin{array}{c}\input{pstex/#2.pstex_t} \end{array}}}{\ensuremath{\begin{array}{c}\input{pstex/#3.pstex_t} \end{array}}}
}

\newcommand{\LD}[1]{
	\settowidth{\LabLength}{\scriptsize \textbf{\ref{#1}}}
	\addtolength{\LabLength}{0.8em}
	\begin{minipage}{\LabLength}
		\scriptsize
		\begin{D}\label{#1}\end{D}
	\end{minipage}
}

\newcommand{\PD}[2]{
	\settowidth{\LabLength}{\scriptsize\textbf{\ref{#1}}}
	\settowidth{\ProcessRefLength}{\scriptsize\ref{#2}}
	\addtolength{\LabLength}{\ProcessRefLength}
	\addtolength{\LabLength}{\ProcessLength}
	\addtolength{\LabLength}{0.8em}
	\begin{minipage}{\LabLength}
		\scriptsize
		\begin{D}\label{#1}$\rightarrow \ref{#2}$\end{D}
	\end{minipage}
}

\newcommand{\CD}[2]{
	\settowidth{\LabLength}{\scriptsize\textbf{\ref{#1}}}
	\settowidth{\CancelRefLength}{\scriptsize$\ref{#2}$}
	\addtolength{\LabLength}{\CancelRefLength}
	\addtolength{\LabLength}{\CancelLength}
	\begin{minipage}{\LabLength}
		\scriptsize
		\begin{DC}\label{#1} \ref{#2} \textbf{\}} \end{DC}
	\end{minipage}
}

\newcommand{\PO}[4][1]{
	\begin{array}{c}
		\PD{#3}{#4}
	\\[#1ex]
		#2
	\end{array}
}

\newcommand{\PDi}[4][1]{
	\PO[#1]{\ensuremath{\begin{array}{c}\input{pstex/#2.pstex_t} \end{array}}}{#3}{#4}
}

\newcommand{\LO}[3][1]{
	\begin{array}{c}
		\LD{#3}
	\\[#1ex]
		#2
	\end{array}
}

\newcommand{\LDi}[3][1]{\LO[#1]{\ensuremath{\begin{array}{c}\input{pstex/#2.pstex_t} \end{array}}}{#3}}

\newcommand{\CDi}[4][1]{
	\begin{array}{c}
		\CD{#3}{#4}
	\\[#1ex]
		\ensuremath{\begin{array}{c}\input{pstex/#2.pstex_t} \end{array}}
	\end{array}
}

\newcommand{\Cancel}[2]{
\begin{cancel}
	Diagram~\ref{#1}  exactly cancels diagram~\ref{#2}.
	\label{cancel:#1}
\end{cancel}
}

\newcommand{\CancelCom}[3]{
\begin{cancel}
Diagram~\ref{#1}  exactly cancels diagram~\ref{#2}#3
\label{cancel:#1}
\end{cancel}
}

\newcommand{\jhep}[3]{{JHEP} #1 (#2) #3}
\newcommand{\NuclPhys}[4]{{Nucl.\ Phys.\ }\textbf{#1 #2} (#3) #4}
\newcommand{\PhysRev}[4]{{Phys.\ Rev.\ }\textbf{#1 #2} (#3) #4}
\newcommand{\IntJModPhys}[4]{{Int.\ J.\ Mod.\ Phys.\ }\textbf{#1 #2} (#3) #4}
\newcommand{\PhysRep}[4]{{Phys.\ Rep.\ }\textbf{#1 #2} (#3) #4}
\newcommand{\PhysRept}[3]{{Phys.\ Rept.\ }\textbf{#1} (#2) #3}

\newcommand{\PhysLett}[4]{{Phys.\ Lett.\ }\textbf{#1 #2} (#3) #4}
\newcommand{\ProgTheorPhys}[3]{{Prog.\ Theor.\ Phys.\ }\textbf{#1} (#2) #3}
\newcommand{\ProgTheorPhysS}[3]{{Prog.\ Theor.\ Phys.\ Suppl.\ }\textbf{#1} (#2) #3}

\newcommand{\CEurJPhys}[3]{{Central Eur.\ J.\ Phys.\ }\textbf{#1} (#2) #3}

\newcommand{\arxiv}[1]{#1}
\newcommand{\hepth}[1]{hep-th/#1}
\newcommand{\hepph}[1]{hep-ph/#1}

\newcommand{\condmat}[1]{cond-mat/#1}

\newcommand{\RevModPhys}[3]{{Rev.\ Mod.\ Phys.\ }\textbf{#1} (#2) #3}

\newcommand{\http}[1]{http://#1}

%% file: pstex/Vertex-S.pstex_t
\begin{picture}(0,0)%
\includegraphics{pstex/Vertex-S.pstex}%
\end{picture}%
\setlength{\unitlength}{3947sp}%
\begingroup\makeatletter\ifx\SetFigFont\undefined%
\gdef\SetFigFont#1#2#3#4#5{%
  \reset@font\fontsize{#1}{#2pt}%
  \fontfamily{#3}\fontseries{#4}\fontshape{#5}%
  \selectfont}%
\fi\endgroup%
\begin{picture}(320,318)(2180,-963)
\put(2291,-859){\makebox(0,0)[lb]{\smash{\SetFigFont{11}{13.2}{\rmdefault}{\mddefault}{\updefault}{\color[rgb]{0,0,0}$S$}%
}}}
\end{picture}

%% file: pstex/Direct.pstex_t
\begin{picture}(0,0)%
\includegraphics{pstex/Direct.pstex}%
\end{picture}%
\setlength{\unitlength}{3947sp}%
\begingroup\makeatletter\ifx\SetFigFont\undefined%
\gdef\SetFigFont#1#2#3#4#5{%
  \reset@font\fontsize{#1}{#2pt}%
  \fontfamily{#3}\fontseries{#4}\fontshape{#5}%
  \selectfont}%
\fi\endgroup%
\begin{picture}(624,477)(2089,-976)
\end{picture}

%% file: pstex/Indirect-2.pstex_t
\begin{picture}(0,0)%
\includegraphics{pstex/Indirect-2.pstex}%
\end{picture}%
\setlength{\unitlength}{3947sp}%
\begingroup\makeatletter\ifx\SetFigFont\undefined%
\gdef\SetFigFont#1#2#3#4#5{%
  \reset@font\fontsize{#1}{#2pt}%
  \fontfamily{#3}\fontseries{#4}\fontshape{#5}%
  \selectfont}%
\fi\endgroup%
\begin{picture}(624,809)(2089,-1279)
\put(2355,-554){\makebox(0,0)[lb]{\smash{\SetFigFont{8}{9.6}{\rmdefault}{\mddefault}{\updefault}{\color[rgb]{0,0,0}$A^2$}%
}}}
\end{picture}

%% file: pstex/Indirect-1.pstex_t
\begin{picture}(0,0)%
\includegraphics{pstex/Indirect-1.pstex}%
\end{picture}%
\setlength{\unitlength}{3947sp}%
\begingroup\makeatletter\ifx\SetFigFont\undefined%
\gdef\SetFigFont#1#2#3#4#5{%
  \reset@font\fontsize{#1}{#2pt}%
  \fontfamily{#3}\fontseries{#4}\fontshape{#5}%
  \selectfont}%
\fi\endgroup%
\begin{picture}(624,809)(2089,-1279)
\put(2355,-554){\makebox(0,0)[lb]{\smash{\SetFigFont{8}{9.6}{\rmdefault}{\mddefault}{\updefault}{\color[rgb]{0,0,0}$A^1$}%
}}}
\end{picture}

%% file: pstex/GR-TP.pstex_t
\begin{picture}(0,0)%
\includegraphics{pstex/GR-TP.pstex}%
\end{picture}%
\setlength{\unitlength}{3947sp}%
\begingroup\makeatletter\ifx\SetFigFont\undefined%
\gdef\SetFigFont#1#2#3#4#5{%
  \reset@font\fontsize{#1}{#2pt}%
  \fontfamily{#3}\fontseries{#4}\fontshape{#5}%
  \selectfont}%
\fi\endgroup%
\begin{picture}(757,318)(1880,-963)
\put(2278,-861){\makebox(0,0)[lb]{\smash{{\SetFigFont{11}{13.2}{\rmdefault}{\mddefault}{\updefault}{\color[rgb]{0,0,0}$S$}%
}}}}
\end{picture}%

%% file: pstex/GR-TP-PF.pstex_t
\begin{picture}(0,0)%
\includegraphics{pstex/GR-TP-PF.pstex}%
\end{picture}%
\setlength{\unitlength}{3947sp}%
\begingroup\makeatletter\ifx\SetFigFont\undefined%
\gdef\SetFigFont#1#2#3#4#5{%
  \reset@font\fontsize{#1}{#2pt}%
  \fontfamily{#3}\fontseries{#4}\fontshape{#5}%
  \selectfont}%
\fi\endgroup%
\begin{picture}(457,414)(2180,-1059)
\put(2278,-861){\makebox(0,0)[lb]{\smash{{\SetFigFont{11}{13.2}{\rmdefault}{\mddefault}{\updefault}{\color[rgb]{0,0,0}$S$}%
}}}}
\end{picture}%

%% file: pstex/GR-TP-PB.pstex_t
\begin{picture}(0,0)%
\includegraphics{pstex/GR-TP-PB.pstex}%
\end{picture}%
\setlength{\unitlength}{3947sp}%
\begingroup\makeatletter\ifx\SetFigFont\undefined%
\gdef\SetFigFont#1#2#3#4#5{%
  \reset@font\fontsize{#1}{#2pt}%
  \fontfamily{#3}\fontseries{#4}\fontshape{#5}%
  \selectfont}%
\fi\endgroup%
\begin{picture}(457,414)(2180,-963)
\put(2278,-861){\makebox(0,0)[lb]{\smash{{\SetFigFont{11}{13.2}{\rmdefault}{\mddefault}{\updefault}{\color[rgb]{0,0,0}$S$}%
}}}}
\end{picture}%

%% file: pstex/Taylor-PFa.pstex_t
\begin{picture}(0,0)%
\includegraphics{pstex/Taylor-PFa.pstex}%
\end{picture}%
\setlength{\unitlength}{3947sp}%
\begingroup\makeatletter\ifx\SetFigFont\undefined%
\gdef\SetFigFont#1#2#3#4#5{%
  \reset@font\fontsize{#1}{#2pt}%
  \fontfamily{#3}\fontseries{#4}\fontshape{#5}%
  \selectfont}%
\fi\endgroup%
\begin{picture}(626,323)(2090,-1016)
\end{picture}

%% file: pstex/Taylor-PFb.pstex_t
\begin{picture}(0,0)%
\includegraphics{pstex/Taylor-PFb.pstex}%
\end{picture}%
\setlength{\unitlength}{3947sp}%
\begingroup\makeatletter\ifx\SetFigFont\undefined%
\gdef\SetFigFont#1#2#3#4#5{%
  \reset@font\fontsize{#1}{#2pt}%
  \fontfamily{#3}\fontseries{#4}\fontshape{#5}%
  \selectfont}%
\fi\endgroup%
\begin{picture}(780,210)(2090,-1016)
\end{picture}

%% file: pstex/Taylor-PBa.pstex_t
\begin{picture}(0,0)%
\includegraphics{pstex/Taylor-PBa.pstex}%
\end{picture}%
\setlength{\unitlength}{3947sp}%
\begingroup\makeatletter\ifx\SetFigFont\undefined%
\gdef\SetFigFont#1#2#3#4#5{%
  \reset@font\fontsize{#1}{#2pt}%
  \fontfamily{#3}\fontseries{#4}\fontshape{#5}%
  \selectfont}%
\fi\endgroup%
\begin{picture}(626,297)(2090,-1016)
\end{picture}

%% file: pstex/Taylor-PBb.pstex_t
\begin{picture}(0,0)%
\includegraphics{pstex/Taylor-PBb.pstex}%
\end{picture}%
\setlength{\unitlength}{3947sp}%
\begingroup\makeatletter\ifx\SetFigFont\undefined%
\gdef\SetFigFont#1#2#3#4#5{%
  \reset@font\fontsize{#1}{#2pt}%
  \fontfamily{#3}\fontseries{#4}\fontshape{#5}%
  \selectfont}%
\fi\endgroup%
\begin{picture}(782,210)(1934,-1016)
\end{picture}

%% file: pstex/DifferentiatedKernel-A.pstex_t
\begin{picture}(0,0)%
\includegraphics{pstex/DifferentiatedKernel-A.pstex}%
\end{picture}%
\setlength{\unitlength}{3947sp}%
\begingroup\makeatletter\ifx\SetFigFont\undefined%
\gdef\SetFigFont#1#2#3#4#5{%
  \reset@font\fontsize{#1}{#2pt}%
  \fontfamily{#3}\fontseries{#4}\fontshape{#5}%
  \selectfont}%
\fi\endgroup%
\begin{picture}(745,453)(2040,-831)
\put(2387,-786){\makebox(0,0)[lb]{\smash{\SetFigFont{11}{13.2}{\rmdefault}{\mddefault}{\updefault}{\color[rgb]{0,0,0}$\DiagDot$}%
}}}
\end{picture}

%% file: pstex/Stub-PF.pstex_t
\begin{picture}(0,0)%
\includegraphics{pstex/Stub-PF.pstex}%
\end{picture}%
\setlength{\unitlength}{3947sp}%
\begingroup\makeatletter\ifx\SetFigFont\undefined%
\gdef\SetFigFont#1#2#3#4#5{%
  \reset@font\fontsize{#1}{#2pt}%
  \fontfamily{#3}\fontseries{#4}\fontshape{#5}%
  \selectfont}%
\fi\endgroup%
\begin{picture}(320,487)(1516,-861)
\end{picture}

%% file: pstex/Stub-PB.pstex_t
\begin{picture}(0,0)%
\includegraphics{pstex/Stub-PB.pstex}%
\end{picture}%
\setlength{\unitlength}{3947sp}%
\begingroup\makeatletter\ifx\SetFigFont\undefined%
\gdef\SetFigFont#1#2#3#4#5{%
  \reset@font\fontsize{#1}{#2pt}%
  \fontfamily{#3}\fontseries{#4}\fontshape{#5}%
  \selectfont}%
\fi\endgroup%
\begin{picture}(320,487)(1516,-861)
\end{picture}

%% file: pstex/Vertex-n_r-B.pstex_t
\begin{picture}(0,0)%
\includegraphics{pstex/Vertex-n_r-B.pstex}%
\end{picture}%
\setlength{\unitlength}{3947sp}%
\begingroup\makeatletter\ifx\SetFigFont\undefined%
\gdef\SetFigFont#1#2#3#4#5{%
  \reset@font\fontsize{#1}{#2pt}%
  \fontfamily{#3}\fontseries{#4}\fontshape{#5}%
  \selectfont}%
\fi\endgroup%
\begin{picture}(320,318)(1776,-676)
\put(1857,-545){\makebox(0,0)[lb]{\smash{\SetFigFont{11}{13.2}{\rmdefault}{\mddefault}{\updefault}{\color[rgb]{0,0,0}$n_r$}%
}}}
\end{picture}

%% file: pstex/K-Delta.pstex_t
\begin{picture}(0,0)%
\includegraphics{pstex/K-Delta.pstex}%
\end{picture}%
\setlength{\unitlength}{3947sp}%
\begingroup\makeatletter\ifx\SetFigFont\undefined%
\gdef\SetFigFont#1#2#3#4#5{%
  \reset@font\fontsize{#1}{#2pt}%
  \fontfamily{#3}\fontseries{#4}\fontshape{#5}%
  \selectfont}%
\fi\endgroup%
\begin{picture}(374,395)(1791,-1006)
\put(1791,-843){\makebox(0,0)[lb]{\smash{\SetFigFont{8}{9.6}{\rmdefault}{\mddefault}{\updefault}{\color[rgb]{0,0,0}$M$}%
}}}
\end{picture}

%% file: pstex/FullGaugeRemainder.pstex_t
\begin{picture}(0,0)%
\includegraphics{pstex/FullGaugeRemainder.pstex}%
\end{picture}%
\setlength{\unitlength}{3947sp}%
\begingroup\makeatletter\ifx\SetFigFont\undefined%
\gdef\SetFigFont#1#2#3#4#5{%
  \reset@font\fontsize{#1}{#2pt}%
  \fontfamily{#3}\fontseries{#4}\fontshape{#5}%
  \selectfont}%
\fi\endgroup%
\begin{picture}(424,395)(2053,-930)
\put(2053,-773){\makebox(0,0)[lb]{\smash{\SetFigFont{8}{9.6}{\rmdefault}{\mddefault}{\updefault}{\color[rgb]{0,0,0}$M$}%
}}}
\end{picture}

%% file: pstex/DecomposedGR.pstex_t
\begin{picture}(0,0)%
\includegraphics{pstex/DecomposedGR.pstex}%
\end{picture}%
\setlength{\unitlength}{3947sp}%
\begingroup\makeatletter\ifx\SetFigFont\undefined%
\gdef\SetFigFont#1#2#3#4#5{%
  \reset@font\fontsize{#1}{#2pt}%
  \fontfamily{#3}\fontseries{#4}\fontshape{#5}%
  \selectfont}%
\fi\endgroup%
\begin{picture}(540,395)(1936,-925)
\put(1936,-776){\makebox(0,0)[lb]{\smash{\SetFigFont{8}{9.6}{\rmdefault}{\mddefault}{\updefault}{\color[rgb]{0,0,0}$M$}%
}}}
\end{picture}

%% file: pstex/GR-TLTP.pstex_t
\begin{picture}(0,0)%
\includegraphics{pstex/GR-TLTP.pstex}%
\end{picture}%
\setlength{\unitlength}{3947sp}%
\begingroup\makeatletter\ifx\SetFigFont\undefined%
\gdef\SetFigFont#1#2#3#4#5{%
  \reset@font\fontsize{#1}{#2pt}%
  \fontfamily{#3}\fontseries{#4}\fontshape{#5}%
  \selectfont}%
\fi\endgroup%
\begin{picture}(757,318)(1880,-963)
\put(2296,-857){\makebox(0,0)[lb]{\smash{\SetFigFont{11}{13.2}{\rmdefault}{\mddefault}{\updefault}{\color[rgb]{0,0,0}$0$}%
}}}
\end{picture}

%% file: pstex/GR-TLTP-EP.pstex_t
\begin{picture}(0,0)%
\includegraphics{pstex/GR-TLTP-EP.pstex}%
\end{picture}%
\setlength{\unitlength}{3947sp}%
\begingroup\makeatletter\ifx\SetFigFont\undefined%
\gdef\SetFigFont#1#2#3#4#5{%
  \reset@font\fontsize{#1}{#2pt}%
  \fontfamily{#3}\fontseries{#4}\fontshape{#5}%
  \selectfont}%
\fi\endgroup%
\begin{picture}(1081,306)(2490,-1356)
\put(2776,-1250){\makebox(0,0)[lb]{\smash{\SetFigFont{11}{13.2}{\rmdefault}{\mddefault}{\updefault}{\color[rgb]{0,0,0}0}%
}}}
\end{picture}

%% file: pstex/k.pstex_t
\begin{picture}(0,0)%
\includegraphics{pstex/k.pstex}%
\end{picture}%
\setlength{\unitlength}{3947sp}%
\begingroup\makeatletter\ifx\SetFigFont\undefined%
\gdef\SetFigFont#1#2#3#4#5{%
  \reset@font\fontsize{#1}{#2pt}%
  \fontfamily{#3}\fontseries{#4}\fontshape{#5}%
  \selectfont}%
\fi\endgroup%
\begin{picture}(174,174)(2239,-821)
\end{picture}

%% file: pstex/kkprk.pstex_t
\begin{picture}(0,0)%
\includegraphics{pstex/kkprk.pstex}%
\end{picture}%
\setlength{\unitlength}{3947sp}%
\begingroup\makeatletter\ifx\SetFigFont\undefined%
\gdef\SetFigFont#1#2#3#4#5{%
  \reset@font\fontsize{#1}{#2pt}%
  \fontfamily{#3}\fontseries{#4}\fontshape{#5}%
  \selectfont}%
\fi\endgroup%
\begin{picture}(499,174)(2239,-820)
\end{picture}

%% file: pstex/GR-relation.pstex_t
\begin{picture}(0,0)%
\includegraphics{pstex/GR-relation.pstex}%
\end{picture}%
\setlength{\unitlength}{3947sp}%
\begingroup\makeatletter\ifx\SetFigFont\undefined%
\gdef\SetFigFont#1#2#3#4#5{%
  \reset@font\fontsize{#1}{#2pt}%
  \fontfamily{#3}\fontseries{#4}\fontshape{#5}%
  \selectfont}%
\fi\endgroup%
\begin{picture}(314,174)(2239,-821)
\end{picture}

%% file: pstex/TLTP-EP-GR.pstex_t
\begin{picture}(0,0)%
\includegraphics{pstex/TLTP-EP-GR.pstex}%
\end{picture}%
\setlength{\unitlength}{3947sp}%
\begingroup\makeatletter\ifx\SetFigFont\undefined%
\gdef\SetFigFont#1#2#3#4#5{%
  \reset@font\fontsize{#1}{#2pt}%
  \fontfamily{#3}\fontseries{#4}\fontshape{#5}%
  \selectfont}%
\fi\endgroup%
\begin{picture}(1059,306)(2512,-1356)
\put(2776,-1250){\makebox(0,0)[lb]{\smash{\SetFigFont{11}{13.2}{\rmdefault}{\mddefault}{\updefault}{\color[rgb]{0,0,0}0}%
}}}
\end{picture}

%% file: pstex/kpr.pstex_t
\begin{picture}(0,0)%
\includegraphics{pstex/kpr.pstex}%
\end{picture}%
\setlength{\unitlength}{3947sp}%
\begingroup\makeatletter\ifx\SetFigFont\undefined%
\gdef\SetFigFont#1#2#3#4#5{%
  \reset@font\fontsize{#1}{#2pt}%
  \fontfamily{#3}\fontseries{#4}\fontshape{#5}%
  \selectfont}%
\fi\endgroup%
\begin{picture}(174,174)(2379,-821)
\end{picture}

%% file: pstex/kprkkpr.pstex_t
\begin{picture}(0,0)%
\includegraphics{pstex/kprkkpr.pstex}%
\end{picture}%
\setlength{\unitlength}{3947sp}%
\begingroup\makeatletter\ifx\SetFigFont\undefined%
\gdef\SetFigFont#1#2#3#4#5{%
  \reset@font\fontsize{#1}{#2pt}%
  \fontfamily{#3}\fontseries{#4}\fontshape{#5}%
  \selectfont}%
\fi\endgroup%
\begin{picture}(508,174)(2379,-817)
\end{picture}

%% file: pstex/EP-GR.pstex_t
\begin{picture}(0,0)%
\includegraphics{pstex/EP-GR.pstex}%
\end{picture}%
\setlength{\unitlength}{3947sp}%
\begingroup\makeatletter\ifx\SetFigFont\undefined%
\gdef\SetFigFont#1#2#3#4#5{%
  \reset@font\fontsize{#1}{#2pt}%
  \fontfamily{#3}\fontseries{#4}\fontshape{#5}%
  \selectfont}%
\fi\endgroup%
\begin{picture}(406,118)(3165,-1234)
\end{picture}

%% file: pstex/EP-GRpr.pstex_t
\begin{picture}(0,0)%
\includegraphics{pstex/EP-GRpr.pstex}%
\end{picture}%
\setlength{\unitlength}{3947sp}%
\begingroup\makeatletter\ifx\SetFigFont\undefined%
\gdef\SetFigFont#1#2#3#4#5{%
  \reset@font\fontsize{#1}{#2pt}%
  \fontfamily{#3}\fontseries{#4}\fontshape{#5}%
  \selectfont}%
\fi\endgroup%
\begin{picture}(628,174)(1926,-821)
\end{picture}

%% file: pstex/GR-PEP.pstex_t
\begin{picture}(0,0)%
\includegraphics{pstex/GR-PEP.pstex}%
\end{picture}%
\setlength{\unitlength}{3947sp}%
\begingroup\makeatletter\ifx\SetFigFont\undefined%
\gdef\SetFigFont#1#2#3#4#5{%
  \reset@font\fontsize{#1}{#2pt}%
  \fontfamily{#3}\fontseries{#4}\fontshape{#5}%
  \selectfont}%
\fi\endgroup%
\begin{picture}(786,174)(2089,-821)
\end{picture}

%% file: pstex/Vertex-1-EP.pstex_t
\begin{picture}(0,0)%
\includegraphics{pstex/Vertex-1-EP.pstex}%
\end{picture}%
\setlength{\unitlength}{3947sp}%
\begingroup\makeatletter\ifx\SetFigFont\undefined%
\gdef\SetFigFont#1#2#3#4#5{%
  \reset@font\fontsize{#1}{#2pt}%
  \fontfamily{#3}\fontseries{#4}\fontshape{#5}%
  \selectfont}%
\fi\endgroup%
\begin{picture}(323,482)(1653,-510)
\put(1771,-395){\makebox(0,0)[lb]{\smash{\SetFigFont{11}{13.2}{\rmdefault}{\mddefault}{\updefault}{\color[rgb]{0,0,0}$1$}%
}}}
\end{picture}

%% file: pstex/Vertex-1.pstex_t
\begin{picture}(0,0)%
\includegraphics{pstex/Vertex-1.pstex}%
\end{picture}%
\setlength{\unitlength}{3947sp}%
\begingroup\makeatletter\ifx\SetFigFont\undefined%
\gdef\SetFigFont#1#2#3#4#5{%
  \reset@font\fontsize{#1}{#2pt}%
  \fontfamily{#3}\fontseries{#4}\fontshape{#5}%
  \selectfont}%
\fi\endgroup%
\begin{picture}(320,318)(1653,-510)
\put(1771,-395){\makebox(0,0)[lb]{\smash{\SetFigFont{11}{13.2}{\rmdefault}{\mddefault}{\updefault}{\color[rgb]{0,0,0}$1$}%
}}}
\end{picture}

%% file: pstex/Vertex-0.pstex_t
\begin{picture}(0,0)%
\includegraphics{pstex/Vertex-0.pstex}%
\end{picture}%
\setlength{\unitlength}{3947sp}%
\begingroup\makeatletter\ifx\SetFigFont\undefined%
\gdef\SetFigFont#1#2#3#4#5{%
  \reset@font\fontsize{#1}{#2pt}%
  \fontfamily{#3}\fontseries{#4}\fontshape{#5}%
  \selectfont}%
\fi\endgroup%
\begin{picture}(320,318)(1653,-510)
\put(1762,-400){\makebox(0,0)[lb]{\smash{\SetFigFont{11}{13.2}{\rmdefault}{\mddefault}{\updefault}{\color[rgb]{0,0,0}$0$}%
}}}
\end{picture}

%% file: pstex/TLTP-EP-GR-B.pstex_t
\begin{picture}(0,0)%
\includegraphics{pstex/TLTP-EP-GR-B.pstex}%
\end{picture}%
\setlength{\unitlength}{3947sp}%
\begingroup\makeatletter\ifx\SetFigFont\undefined%
\gdef\SetFigFont#1#2#3#4#5{%
  \reset@font\fontsize{#1}{#2pt}%
  \fontfamily{#3}\fontseries{#4}\fontshape{#5}%
  \selectfont}%
\fi\endgroup%
\begin{picture}(439,594)(1649,144)
\put(1751,332){\makebox(0,0)[lb]{\smash{\SetFigFont{8}{9.6}{\rmdefault}{\mddefault}{\updefault}{\color[rgb]{0,0,0}$0^2$}%
}}}
\end{picture}

%% file: pstex/TLTP-GR-PEP.pstex_t
\begin{picture}(0,0)%
\includegraphics{pstex/TLTP-GR-PEP.pstex}%
\end{picture}%
\setlength{\unitlength}{3947sp}%
\begingroup\makeatletter\ifx\SetFigFont\undefined%
\gdef\SetFigFont#1#2#3#4#5{%
  \reset@font\fontsize{#1}{#2pt}%
  \fontfamily{#3}\fontseries{#4}\fontshape{#5}%
  \selectfont}%
\fi\endgroup%
\begin{picture}(438,603)(1649,163)
\put(1751,332){\makebox(0,0)[lb]{\smash{\SetFigFont{8}{9.6}{\rmdefault}{\mddefault}{\updefault}{\color[rgb]{0,0,0}$0^2$}%
}}}
\end{picture}

%% file: pstex/GR-LdL.pstex_t
\begin{picture}(0,0)%
\includegraphics{pstex/GR-LdL.pstex}%
\end{picture}%
\setlength{\unitlength}{3947sp}%
\begingroup\makeatletter\ifx\SetFigFont\undefined%
\gdef\SetFigFont#1#2#3#4#5{%
  \reset@font\fontsize{#1}{#2pt}%
  \fontfamily{#3}\fontseries{#4}\fontshape{#5}%
  \selectfont}%
\fi\endgroup%
\begin{picture}(205,238)(1748,126)
\put(1914,157){\makebox(0,0)[lb]{\smash{\SetFigFont{8}{9.6}{\rmdefault}{\mddefault}{\updefault}{\color[rgb]{0,0,0}$\bullet$}%
}}}
\end{picture}

%% file: pstex/GR-GR-LdL.pstex_t
\begin{picture}(0,0)%
\includegraphics{pstex/GR-GR-LdL.pstex}%
\end{picture}%
\setlength{\unitlength}{3947sp}%
\begingroup\makeatletter\ifx\SetFigFont\undefined%
\gdef\SetFigFont#1#2#3#4#5{%
  \reset@font\fontsize{#1}{#2pt}%
  \fontfamily{#3}\fontseries{#4}\fontshape{#5}%
  \selectfont}%
\fi\endgroup%
\begin{picture}(217,349)(1748,126)
\put(1914,157){\makebox(0,0)[lb]{\smash{\SetFigFont{8}{9.6}{\rmdefault}{\mddefault}{\updefault}{\color[rgb]{0,0,0}$\bullet$}%
}}}
\end{picture}

%% file: pstex/Vertex-1-EP-WBT.pstex_t
\begin{picture}(0,0)%
\includegraphics{pstex/Vertex-1-EP-WBT.pstex}%
\end{picture}%
\setlength{\unitlength}{3947sp}%
\begingroup\makeatletter\ifx\SetFigFont\undefined%
\gdef\SetFigFont#1#2#3#4#5{%
  \reset@font\fontsize{#1}{#2pt}%
  \fontfamily{#3}\fontseries{#4}\fontshape{#5}%
  \selectfont}%
\fi\endgroup%
\begin{picture}(320,782)(1653,-510)
\put(1770,-409){\makebox(0,0)[lb]{\smash{\SetFigFont{11}{13.2}{\rmdefault}{\mddefault}{\updefault}{\color[rgb]{0,0,0}$1$}%
}}}
\end{picture}

%% file: pstex/Struc-AR1-A+B.pstex_t
\begin{picture}(0,0)%
\includegraphics{pstex/Struc-AR1-A+B.pstex}%
\end{picture}%
\setlength{\unitlength}{3947sp}%
\begingroup\makeatletter\ifx\SetFigFont\undefined%
\gdef\SetFigFont#1#2#3#4#5{%
  \reset@font\fontsize{#1}{#2pt}%
  \fontfamily{#3}\fontseries{#4}\fontshape{#5}%
  \selectfont}%
\fi\endgroup%
\begin{picture}(476,671)(412,340)
\put(708,634){\makebox(0,0)[lb]{\smash{\SetFigFont{8}{9.6}{\rmdefault}{\mddefault}{\updefault}{\color[rgb]{0,0,0}$0^2$}%
}}}
\end{picture}

%% file: pstex/Struc-AR1-A+B-b.pstex_t
\begin{picture}(0,0)%
\includegraphics{pstex/Struc-AR1-A+B-b.pstex}%
\end{picture}%
\setlength{\unitlength}{3947sp}%
\begingroup\makeatletter\ifx\SetFigFont\undefined%
\gdef\SetFigFont#1#2#3#4#5{%
  \reset@font\fontsize{#1}{#2pt}%
  \fontfamily{#3}\fontseries{#4}\fontshape{#5}%
  \selectfont}%
\fi\endgroup%
\begin{picture}(476,671)(854,340)
\put(909,620){\makebox(0,0)[lb]{\smash{{\SetFigFont{8}{9.6}{\rmdefault}{\mddefault}{\updefault}{\color[rgb]{0,0,0}$0^2$}%
}}}}
\end{picture}%

%% file: pstex/TLTP-EP-d.pstex_t
\begin{picture}(0,0)%
\includegraphics{pstex/TLTP-EP-d.pstex}%
\end{picture}%
\setlength{\unitlength}{3947sp}%
\begingroup\makeatletter\ifx\SetFigFont\undefined%
\gdef\SetFigFont#1#2#3#4#5{%
  \reset@font\fontsize{#1}{#2pt}%
  \fontfamily{#3}\fontseries{#4}\fontshape{#5}%
  \selectfont}%
\fi\endgroup%
\begin{picture}(1057,381)(1244,-855)
\put(1428,-702){\makebox(0,0)[lb]{\smash{\SetFigFont{8}{9.6}{\rmdefault}{\mddefault}{\updefault}{\color[rgb]{0,0,0}$0^2$}%
}}}
\end{picture}

%% file: pstex/TLTPd-EP.pstex_t
\begin{picture}(0,0)%
\includegraphics{pstex/TLTPd-EP.pstex}%
\end{picture}%
\setlength{\unitlength}{3947sp}%
\begingroup\makeatletter\ifx\SetFigFont\undefined%
\gdef\SetFigFont#1#2#3#4#5{%
  \reset@font\fontsize{#1}{#2pt}%
  \fontfamily{#3}\fontseries{#4}\fontshape{#5}%
  \selectfont}%
\fi\endgroup%
\begin{picture}(662,357)(1290,-820)
\put(1428,-702){\makebox(0,0)[lb]{\smash{\SetFigFont{8}{9.6}{\rmdefault}{\mddefault}{\updefault}{\color[rgb]{0,0,0}$0^2$}%
}}}
\end{picture}

%% file: pstex/TLTP-EPd.pstex_t
\begin{picture}(0,0)%
\includegraphics{pstex/TLTP-EPd.pstex}%
\end{picture}%
\setlength{\unitlength}{3947sp}%
\begingroup\makeatletter\ifx\SetFigFont\undefined%
\gdef\SetFigFont#1#2#3#4#5{%
  \reset@font\fontsize{#1}{#2pt}%
  \fontfamily{#3}\fontseries{#4}\fontshape{#5}%
  \selectfont}%
\fi\endgroup%
\begin{picture}(662,436)(1290,-820)
\put(1428,-702){\makebox(0,0)[lb]{\smash{\SetFigFont{8}{9.6}{\rmdefault}{\mddefault}{\updefault}{\color[rgb]{0,0,0}$0^2$}%
}}}
\end{picture}

%% file: pstex/GRd.pstex_t
\begin{picture}(0,0)%
\includegraphics{pstex/GRd.pstex}%
\end{picture}%
\setlength{\unitlength}{3947sp}%
\begingroup\makeatletter\ifx\SetFigFont\undefined%
\gdef\SetFigFont#1#2#3#4#5{%
  \reset@font\fontsize{#1}{#2pt}%
  \fontfamily{#3}\fontseries{#4}\fontshape{#5}%
  \selectfont}%
\fi\endgroup%
\begin{picture}(240,387)(1795,-724)
\end{picture}

%% file: pstex/DiffGR.pstex_t
\begin{picture}(0,0)%
\includegraphics{pstex/DiffGR.pstex}%
\end{picture}%
\setlength{\unitlength}{3947sp}%
\begingroup\makeatletter\ifx\SetFigFont\undefined%
\gdef\SetFigFont#1#2#3#4#5{%
  \reset@font\fontsize{#1}{#2pt}%
  \fontfamily{#3}\fontseries{#4}\fontshape{#5}%
  \selectfont}%
\fi\endgroup%
\begin{picture}(240,398)(266,-72)
\end{picture}

%% file: pstex/DiffGRk.pstex_t
\begin{picture}(0,0)%
\includegraphics{pstex/DiffGRk.pstex}%
\end{picture}%
\setlength{\unitlength}{3947sp}%
\begingroup\makeatletter\ifx\SetFigFont\undefined%
\gdef\SetFigFont#1#2#3#4#5{%
  \reset@font\fontsize{#1}{#2pt}%
  \fontfamily{#3}\fontseries{#4}\fontshape{#5}%
  \selectfont}%
\fi\endgroup%
\begin{picture}(324,398)(266,-72)
\end{picture}

%% file: pstex/DiffGRkpr.pstex_t
\begin{picture}(0,0)%
\includegraphics{pstex/DiffGRkpr.pstex}%
\end{picture}%
\setlength{\unitlength}{3947sp}%
\begingroup\makeatletter\ifx\SetFigFont\undefined%
\gdef\SetFigFont#1#2#3#4#5{%
  \reset@font\fontsize{#1}{#2pt}%
  \fontfamily{#3}\fontseries{#4}\fontshape{#5}%
  \selectfont}%
\fi\endgroup%
\begin{picture}(286,398)(360,-76)
\end{picture}

%% file: pstex/GR-W-dTLTP-EP.pstex_t
\begin{picture}(0,0)%
\includegraphics{pstex/GR-W-dTLTP-EP.pstex}%
\end{picture}%
\setlength{\unitlength}{3947sp}%
\begingroup\makeatletter\ifx\SetFigFont\undefined%
\gdef\SetFigFont#1#2#3#4#5{%
  \reset@font\fontsize{#1}{#2pt}%
  \fontfamily{#3}\fontseries{#4}\fontshape{#5}%
  \selectfont}%
\fi\endgroup%
\begin{picture}(869,371)(1219,5)
\put(1615,194){\makebox(0,0)[lb]{\smash{\SetFigFont{11}{13.2}{\rmdefault}{\mddefault}{\updefault}{\color[rgb]{0,0,0}0}%
}}}
\end{picture}

%% file: pstex/GR-W-TLTP-d-EP.pstex_t
\begin{picture}(0,0)%
\includegraphics{pstex/GR-W-TLTP-d-EP.pstex}%
\end{picture}%
\setlength{\unitlength}{3947sp}%
\begingroup\makeatletter\ifx\SetFigFont\undefined%
\gdef\SetFigFont#1#2#3#4#5{%
  \reset@font\fontsize{#1}{#2pt}%
  \fontfamily{#3}\fontseries{#4}\fontshape{#5}%
  \selectfont}%
\fi\endgroup%
\begin{picture}(1187,369)(901,86)
\put(1615,194){\makebox(0,0)[lb]{\smash{\SetFigFont{11}{13.2}{\rmdefault}{\mddefault}{\updefault}{\color[rgb]{0,0,0}0}%
}}}
\end{picture}

%% file: pstex/GR-dW-TLTP-EP.pstex_t
\begin{picture}(0,0)%
\includegraphics{pstex/GR-dW-TLTP-EP.pstex}%
\end{picture}%
\setlength{\unitlength}{3947sp}%
\begingroup\makeatletter\ifx\SetFigFont\undefined%
\gdef\SetFigFont#1#2#3#4#5{%
  \reset@font\fontsize{#1}{#2pt}%
  \fontfamily{#3}\fontseries{#4}\fontshape{#5}%
  \selectfont}%
\fi\endgroup%
\begin{picture}(869,386)(1219,-10)
\put(1615,194){\makebox(0,0)[lb]{\smash{\SetFigFont{11}{13.2}{\rmdefault}{\mddefault}{\updefault}{\color[rgb]{0,0,0}0}%
}}}
\end{picture}

%% file: pstex/dGR-W-TLTP-EP.pstex_t
\begin{picture}(0,0)%
\includegraphics{pstex/dGR-W-TLTP-EP.pstex}%
\end{picture}%
\setlength{\unitlength}{3947sp}%
\begingroup\makeatletter\ifx\SetFigFont\undefined%
\gdef\SetFigFont#1#2#3#4#5{%
  \reset@font\fontsize{#1}{#2pt}%
  \fontfamily{#3}\fontseries{#4}\fontshape{#5}%
  \selectfont}%
\fi\endgroup%
\begin{picture}(919,431)(1169,-55)
\put(1615,194){\makebox(0,0)[lb]{\smash{\SetFigFont{11}{13.2}{\rmdefault}{\mddefault}{\updefault}{\color[rgb]{0,0,0}0}%
}}}
\end{picture}

%% file: pstex/LdLdGR-EP.pstex_t
\begin{picture}(0,0)%
\includegraphics{pstex/LdLdGR-EP.pstex}%
\end{picture}%
\setlength{\unitlength}{3947sp}%
\begingroup\makeatletter\ifx\SetFigFont\undefined%
\gdef\SetFigFont#1#2#3#4#5{%
  \reset@font\fontsize{#1}{#2pt}%
  \fontfamily{#3}\fontseries{#4}\fontshape{#5}%
  \selectfont}%
\fi\endgroup%
\begin{picture}(456,467)(1174,-48)
\end{picture}

%% file: pstex/GR-dW.pstex_t
\begin{picture}(0,0)%
\includegraphics{pstex/GR-dW.pstex}%
\end{picture}%
\setlength{\unitlength}{3947sp}%
\begingroup\makeatletter\ifx\SetFigFont\undefined%
\gdef\SetFigFont#1#2#3#4#5{%
  \reset@font\fontsize{#1}{#2pt}%
  \fontfamily{#3}\fontseries{#4}\fontshape{#5}%
  \selectfont}%
\fi\endgroup%
\begin{picture}(404,370)(1219,-8)
\end{picture}

%% file: pstex/GR-dW-GR.pstex_t
\begin{picture}(0,0)%
\includegraphics{pstex/GR-dW-GR.pstex}%
\end{picture}%
\setlength{\unitlength}{3947sp}%
\begingroup\makeatletter\ifx\SetFigFont\undefined%
\gdef\SetFigFont#1#2#3#4#5{%
  \reset@font\fontsize{#1}{#2pt}%
  \fontfamily{#3}\fontseries{#4}\fontshape{#5}%
  \selectfont}%
\fi\endgroup%
\begin{picture}(467,370)(1219,-8)
\end{picture}

%% file: pstex/dGR-W.pstex_t
\begin{picture}(0,0)%
\includegraphics{pstex/dGR-W.pstex}%
\end{picture}%
\setlength{\unitlength}{3947sp}%
\begingroup\makeatletter\ifx\SetFigFont\undefined%
\gdef\SetFigFont#1#2#3#4#5{%
  \reset@font\fontsize{#1}{#2pt}%
  \fontfamily{#3}\fontseries{#4}\fontshape{#5}%
  \selectfont}%
\fi\endgroup%
\begin{picture}(456,396)(1174,-48)
\end{picture}

%% file: pstex/dGR-W-GR.pstex_t
\begin{picture}(0,0)%
\includegraphics{pstex/dGR-W-GR.pstex}%
\end{picture}%
\setlength{\unitlength}{3947sp}%
\begingroup\makeatletter\ifx\SetFigFont\undefined%
\gdef\SetFigFont#1#2#3#4#5{%
  \reset@font\fontsize{#1}{#2pt}%
  \fontfamily{#3}\fontseries{#4}\fontshape{#5}%
  \selectfont}%
\fi\endgroup%
\begin{picture}(522,396)(1174,-48)
\end{picture}